\documentclass[11pt]{article}

\newcommand{\thistitle}{Extended Phase Space in General Gauge Theories}

\usepackage{amsmath,amssymb,amsfonts,bbm,bm}
\usepackage{slashed}

\usepackage{tocloft} 
\usepackage[makeroom]{cancel}

\usepackage[nosort]{cite} 
\usepackage{graphicx,color}
\usepackage[colorlinks=true, linkcolor=blue, citecolor=blue, linktoc=all]{hyperref}

\usepackage[T1]{fontenc}

\usepackage[usenames,dvipsnames]{xcolor}

\usepackage{tikz-cd}
\usepackage{pict2e}
\usepackage{physics}
\usepackage{comment} 
\usepackage[makeroom]{cancel}

\newcommand{\beq}{\begin{eqnarray}}
\newcommand{\eeq}{\end{eqnarray}}
\newcommand{\beqn}{\begin{eqnarray}}
\newcommand{\eeqn}{\end{eqnarray}}
\newcommand{\bea}{\begin{eqnarray}}
\newcommand{\eea}{\end{eqnarray}}
\newcommand{\be}{\begin{equation}}
\newcommand{\ee}{\end{equation}}

\newcommand{\un}[1]{\underline{#1}}
\def\pa{\partial}
\usepackage[normalem]{ulem}

\newcommand{\hlt}[1]{{\color{Emerald}{\em #1}}}

\newcommand{\oseq}{\,\,\hat{=}\,\,}

\newcommand{\uiuc}[1]{
	\centerline{
		\begin{minipage}[c]{0.7\textwidth}
			\begin{center}
			${}^{#1}$ Illinois Center for Advanced Studies of the Universe \& Department of Physics,\\ 
			University of Illinois, 1110 West Green St., Urbana IL 61801, U.S.A.
			\end{center}
		\end{minipage}
		}
	}

\usepackage{mathrsfs,mathtools}
\renewcommand\mathbb[1]{\mathbbm{#1}}
\newcommand\mathbbb[1]{\boldsymbol{\mathbb{#1}}}

\newcommand{\Lagr}{\mathfrak{L}} 

\newcommand{\td}{{\rm d}}

\newcommand{\fsLie}[1]{\text{L}_{\mathbb{V}_{#1}}}
\newcommand{\fscon}[1]{\text{I}_{\mathbb{V}_{#1}}}

\newcommand{\hatd}{\hat{\td}}
\newcommand{\hatLie}[1]{\hat{\cal L}_{#1}}
\newcommand{\hatcon}[1]{\hat{i}_{#1}}

\newcommand{\mX}{\un{\mathfrak{X}}}
\newcommand{\mY}{\un{\mathfrak{Y}}}

\newcommand{\mg}{\mathfrak{g}}

\newcommand{\Aconn}[1]{\phi_{#1}}

\newcommand{\morphalg}{\mathbbb{G}}
\newcommand{\pbalg}{\mathbbb{P}}
\newcommand{\Tpbalg}{\text{T}\pbalg}
\newcommand{\spalg}{\mathbbb{M}}

\newcommand{\fldspalg}{\mathbbb{F}}
\newcommand{\algalg}{\mathbbb{A}}
\newcommand{\ellalg}{\mathbbb{L}}

\newcommand{\alghatd}{\hat{\delta}}
\newcommand{\alghatdTM}{\delta}

\newcommand{\algvec}[1]{\mathbb{V}_{#1}}
\newcommand{\algLie}[1]{\hat{\text{L}}_{#1}}
\newcommand{\algcon}[1]{\hat{\text{I}}_{{#1}}}
\newcommand{\algchg}[1]{\text{H}_{#1}}
\newcommand{\algchgdens}[1]{\text{Q}_{#1}}
\newcommand{\algcurrdens}[1]{\text{J}_{#1}}
\newcommand{\algEnd}[1]{\text {v}_{#1}}

\newcommand{\algsec}[1]{\un{\bm{\mathfrak{#1}}}}

\newcommand{\alginjo}{\mathbbb{j}}
\newcommand{\algrho}{\mathbbb{r}}

\newcommand{\MCb}{\bm{\varpi}}

\newcommand{\ellsec}[1]{\bm{#1}}

\makeatletter
\DeclareRobustCommand{\loplus}{\mathbin{\mathpalette\dog@lsemi{+}}}
\DeclareRobustCommand{\lotimes}{\mathbin{\mathpalette\dog@lsemi{\times}}}
\DeclareRobustCommand{\roplus}{\mathbin{\mathpalette\dog@rsemi{+}}}
\DeclareRobustCommand{\rotimes}{\mathbin{\mathpalette\dog@rsemi{\times}}}

\newcommand{\dog@rsemi}[2]{\dog@semi{#1}{#2}{-90,90}}
\newcommand{\dog@lsemi}[2]{\dog@semi{#1}{#2}{270,90}}
\newcommand{\dog@semi}[3]{%
  \begingroup
  \sbox\z@{$\m@th#1#2$}%
  \setlength{\unitlength}{\dimexpr\ht\z@+\dp\z@\relax}%
  \makebox[\wd\z@]{\raisebox{-\dp\z@}{%
    \begin{picture}(1,1)
    \linethickness{\variable@rule{#1}}
    \roundcap
    \put(0.5,0.5){\makebox(0,0){\raisebox{\dp\z@}{$\m@th#1#2$}}}
    \put(0.5,0.5){\arc[#3]{0.5}}
    \end{picture}%
  }}%
  \endgroup
}
\newcommand{\variable@rule}[1]{%
  \fontdimen8  
  \ifx#1\displaystyle\textfont3\else
    \ifx#1\textstyle\textfont3\else
      \ifx#1\scriptstyle\scriptfont3\else
        \scriptscriptfont3\relax
  \fi\fi\fi
}
\DeclareRobustCommand{\loplus}{\mathbin{\mathpalette\dog@lsemi{+}}}

\newcommand{\gmu}{\mu}
\newcommand{\gnu}{\nu}
\newcommand{\disc}[1]{{\color{red}#1}}
\newcommand{\tD}{{\rm D}}

\begin{document}

\title{\thistitle}
\author{
	Marc S. Klinger, 
	Robert G. Leigh, 
	and 
	Pin-Chun Pai 
	\\
	\\
	{\small \emph{\uiuc{}}}
	\\
	}
\date{}
\maketitle
\vspace{-0.5cm}
\begin{abstract}
\vspace{0.3cm}
In a recent paper, it was shown that in diffeomorphism-invariant theories, Noether charges associated with a given codimension-2 surface become integrable if one introduces an extended phase space.  
In this paper we extend the notion of extended phase space to all gauge theories with arbitrary combinations of internal and spacetime local symmetries. We formulate this in terms of a corresponding Atiyah Lie algebroid, a geometric object derived from a principal bundle which features internal symmetries and diffeomorphisms on an equal footing. In this language, gauge transformations are understood as \emph{morphisms} between Atiyah Lie algebroids that preserve the geometric structures encoded therein. The extended configuration space of a gauge theory can subsequently be understood as the space of pairs $(\varphi, \Phi)$, where $\varphi$ is a Lie algebroid morphism and $\Phi$ is a field configuration in the non-extended sense. Starting from this data, we outline a very powerful, manifestly geometric approach to the extended phase space. Using this approach, we find that the action of the group of gauge transformations and diffeomorphisms on the symplectic geometry of \emph{any} covariant theory is integrable.  We motivate our construction by carefully examining the need for extended phase space in Chern-Simons gauge theories and display its usefulness by re-computing the charge algebra. We also describe the implementation of the configuration algebroid in Einstein-Yang-Mills theories.
\end{abstract}


\setcounter{footnote}{0}
\renewcommand{\thefootnote}{\arabic{footnote}}
\newpage

\section{Introduction}

In \cite{Ciambelli:2021nmv} it was demonstrated how the integrability of the Noether charge algebra arising in a diffeomorphism invariant theory can be reconciled by \emph{extending} the phase space of the theory to include not only the field degrees of freedom, but also degrees of freedom associated with the embedding of subregions in bulk spacetime. In this way the idea of an ``extended phase space'' appears as a natural remedy to the ills that befall the conventional covariant phase space formalism \cite{Crnkovic:1986ex,Crnkovic:1987tz,Compere:2018aar,cmp/1103899589,GAWEDZKI1972307,cmp/1103858807} in cases where gauge symmetries, like diffeomorphisms, have non-zero charges.\footnote{We should note that there are other approaches besides the extended phase space for  obtaining integrable charge algebras in gauge theories such as \cite{Newman:1961qr,barnich2012note,Barnich:2016lyg,barnich2020bms,Rejzner:2020xid}. However, these approaches are algebraic as opposed to geometric, and typically proceed by incorporating new constraints and/or modifying the charge algebra, as opposed to extending the degrees of freedom included in the theory.} The importance of corners (codimension-2 surfaces) in this story is a natural consequence of Noether's second theorem which dictates that the charges associated with gauge symmetries  localize on such codimension-$2$ subregions in bulk spacetime \cite{noether1971invariant} once constraints are satisfied. In this construction, it is precisely the embedding map that carries the degrees of freedom that render a corner, even at finite distance, physically relevant.

In contemporary literature, the study of the conserved quantities associated with gauge symmetries is often intimately related with the idea of ``asymptotic symmetries,'' highlighting the fact that the physical impact of gauge symmetries is generically realized in the subregions where the surface integrals that define their conserved charges are evaluated. For an up to date introduction and review of asymptotic symmetries see \cite{Ciambelli:2022vot,Freidel:2023bnj} and the references therein. Ultimately, it has become clear that the presence of an extended phase space which properly encodes the degrees of freedom associated with gauge charges is a necessary feature in the complete symplectic formulation of any gauge theory. Thus, it is a crucial goal to determine a setting that is up to the task of geometrizing this data in a natural way.

This paper is a direct follow up to \cite{Ciambelli:2021nmv} in which we aim to accomplish the aforementioned goal by formalizing and expanding the approach of that paper to accommodate theories that  possess internal gauge symmetries in addition, possibly, to diffeomorphism invariance. The resulting generalization of the extended phase space is a new geometric construction that successfully encodes the degrees of freedom associated with field configurations, diffeomorphisms, and gauge transformations. We refer to this construction as the \emph{configuration algebroid} for reasons that will become clear. Although it may seem unusual to include diffeomorphisms in the context of ordinary gauge theories since they are usually considered on a fixed background geometry, it is, again, well established that in the presence of subregions, degrees of freedom which were once \emph{pure gauge} may become physical \cite{Ciambelli:2021nmv,Balachandran:1994up,Carlip:1994gy,Carlip:1996yb,Balachandran:1995qa,REGGE1974286,Ciambelli:2021vnn,Ciambelli:2022cfr,Donnelly:2016auv,Donnelly:2020xgu,Geiller:2017xad,Freidel:2020xyx,Freidel:2020svx,Freidel:2020ayo,Freidel:2021dxw,Chandrasekaran:2021hxc,Donnelly:2022kfs}. This is particularly clear in the entanglement properties of gauge theories \cite{wald1993black,iyer1994some,Fliss:2017wop,Banados:1992wn,Strominger:1997eq,Donnelly:2011hn,Donnelly:2014fua,Donnelly:2014gva,Donnelly:2015hxa,Das:2015oha,Wen:2016snr,Carlip:2017xne,Chen:2020nyh,Speranza:2017gxd,Geiller:2019bti,faulkner2016modular,faulkner2016shape,
balasubramanian2017multi,balasubramanian2018entanglement,Chandrasekaran:2020wwn}.\footnote{In the time since the original posting of this note, further work has been completed which explores the role of extended phase space in the operator algebraic formulation of entanglement entropy for subregions in QFTs \cite{Klinger:2023tgi}.} Thus a significant benefit of the formalism introduced in this paper is that it is applicable equally to dynamical geometry as well as backgrounds, with the ability to focus on subregions a byproduct. 

A crucial ingredient in constructing the configuration algebroid is the concept of an \emph{Atiyah Lie algebroid}, which is closely related to the perhaps more familiar concept of a principal bundle \cite{pradines1967theorie,mackenzie2005general,crainic2003integrability}. The Lie algebroid, in general, has received much attention in the mathematics community as an efficient tool for studying integrability in the context of bracket algebras, cohomology for generalized exterior algebras, and symplectic geometry \cite{crainic2003integrability,fernandes2002lie,
roytenberg2002structure,roytenberg1999courant,roytenberg2007aksz}. In recent years, the Lie algebroid has begun to be appreciated in the physics literature, especially as a device for properly encoding the geometry of gauge theories \cite{Fournel:2012uv,Ciambelli:2021ujl,Jia:2023tki,Blohmann:2010jd,LAZZARINI2012387,Carow-Watamura:2016lob,Kotov:2016lpx,ATTARD2020103541,Strobl:2004im,BOJOWALD2005400,Mayer:2009wf}. The most significant feature of the Atiyah Lie algebroid for the purpose of this note is that it treats diffeomorphisms and gauge transformations on an equal footing, and thereby offers a natural language for encoding the degrees of freedom associated with local symmetries in a consistent, and manifestly geometric fashion. We stress this point as a significant departure from standard techniques for analyzing the charge algebra of a theory with internal gauge symmetries in which, typically, one regards diffeomorphisms as a separate issue. By using the Lie algebroid formalism, we arrive at a natural setting in which gauge and gravity are both always at play. For a complete introduction to Lie algebroids see \cite{mackenzie2005general}. Those looking for a more physics-oriented introduction are referred to \cite{Ciambelli:2021ujl}. 

The main result of the present paper is to show that the action of the set of gauge transformations and diffeomorphisms on symplectic geometries defined by Lagrangian theories is both \emph{integrable} and \emph{non-centrally extended} when regarded from the perspective of a suitably constructed extended phase space. Integrability means that each local symmetry generator can be associated with a section of the extended phase space which is {Hamiltonian} in the standard sense -- that is a section whose contraction with the (pre)symplectic form is a total variation. A different way of phrasing the problem of integrability is in terms of the existence of a moment map \cite{souriau1966quantification,souriau1997structure,guillemin1984normal,donaldson1999moment} from the symmetry algebra into the space of charges for the theory. Being non-centrally extended goes beyond integrability, and refers to a circumstance in which the aforementioned moment map is upgraded to an \emph{equivariant} moment map, or alternatively a moment \emph{morphism}. A moment map becomes a moment morphism precisely if it is a morphism between the algebra of the symmetry generators and the Poisson algebra of their associated charges \cite{Kostant:1969zz,kostant1970quantization,atiyah1984moment,blohmann2018hamiltonian}.  The existence of a moment morphism is therefore equivalent to the existence of a non centrally extended Poisson algebra. (See, for example, \cite{brown1986central,barnich2002covariant} for a review of central charges in the context of the covariant phase space formalism.) Crucially, our main result uses neither the equations of motion specific to a given theory, nor constraints tailored to any particular theory. Thus, we conclude that the fact that covariant Lagrangians give rise to \emph{Hamiltonian actions}\footnote{The action of a symmetry group on a symplectic geometry is referred to as Hamiltonian if it is both integrable and non-centrally extended \cite{blohmann2018hamiltonian}.} in the context of the extended phase space of the configuration algebroid is a geometric property of the configuration algebroid itself, rather than a feature of a given theory, whether classical or quantum.  

The organization of this paper is as follows. To enter the arena in which extended phase space is required in gauge theories, it is useful to consider Chern-Simons theories, as they have properties in the gauge sector that are quite similar to other diffeomorphism-invariant theories. In Section \ref{sec:conventional} then, we introduce the need for an extended phase space by demonstrating the failure of integrability in $3$-dimensional Chern-Simons theory when analyzed from the perspective of the conventional covariant phase space formalism. In Section \ref{sec:Primer}, we recall some important aspects of Atiyah Lie algebroids that are central to our construction. In this case, an extension is required both for diffeomorphisms as well as gauge transformations. We emphasize the role of Lie algebroid morphisms as a conceptual replacement for the notion of gauge transformations.  In Section \ref{sec:confalg}, we discuss the space of fields in a gauge theory, and rather than descending to the quotient by the group of morphisms, we pass to a principal bundle $\pbalg$ over the space of fields that we show is the appropriate general setting for extended phase space. We refer to this principal bundle as the extended configuration space. A closely related structure is the \hlt{configuration algebroid} $\algalg$ which properly organizes all of the ingredients of the extended phase space of a gauge theory.   We apply the configuration algebroid to the analysis of the extended phase space in Section \ref{sec: General Proof}, and subsequently prove that the action of the local symmetries on the resulting symplectic geometry is Hamiltonian for the most general case. As was true in \cite{Ciambelli:2021nmv}, these results do not rely on setting some flux to zero, but rather are accomplished generically \emph{off-shell}. Finally, we revisit Chern-Simons theory in Section \ref{Examples} in order to display how our general machinery applies in that case. In Appendix \ref{AppNot}, we collect our notation conventions. In Appendix \ref{EYMApp}, we show how our construction is applied in Einstein-Yang-Mills theories. Finally in Appendix \ref{MorphEmb}, we explain how, as was mentioned in several places in the main body of the paper, that the embedding map that featured in  \cite{Ciambelli:2021nmv} is replaced in the context of Lie algebroids by a Lie algebroid morphism. Thus, in general one considers the definition of a Noether charge as including an embedding map of a corner in addition to a choice of gauge.

\section{Conventional Approach to the Extended Phase Space}
\label{sec:conventional}

In standard notation, a Chern-Simons Lagrangian is written 
\beq\label{CS Spacetime Lagrangian}
\Lagr=\frac{k}{4\pi} tr\Big(A\wedge \td A+\frac13 A\wedge [A\wedge A]\Big)
\eeq
where $A$ denotes a gauge field, interpreted as a 1-form on a 3-dimensional space-time $M$ valued in a Lie algebra $\mg$. 
We then have
\beq
\delta \Lagr
=\frac{k}{2\pi} tr\Big(\delta A\wedge F+\frac12d\Big[\delta A\wedge A\Big]\Big)
\eeq
from which we deduce the presymplectic 1-form
\beq
\theta=\frac{k}{4\pi}tr\, \delta A\wedge A.
\eeq
This is a 2-form in space-time and a 1-form on the space of fields, so we refer to it as a $(2,1)$-form.
We note that the action is diffeomorphism-invariant and gauge-invariant, up to boundary terms. That is, denoting an infinitesimal gauge transformation by $\gmu\in\mg$ and\footnote{For notation conventions used in this paper, see Appendix \ref{AppNot}.}
\beq \label{Gauge transform of A}
\delta_{\gmu} A=\fsLie{\gmu}A=\td\gmu+[A,\gmu]=\tD\gmu
\eeq
we have\footnote{From this point onward, we will use the notation $\hat{=}$ to denote equality \emph{on-shell}.}
\beqn \label{Failure of equivariance}
\delta_{\gmu}\Lagr&=&\fsLie{\gmu}\Lagr\oseq
\td\Big[\fscon{\gmu}\theta\Big]\;=\; \td\Big[\frac{k}{4\pi}\tD\gmu\wedge A\Big]\oseq\td\Big[\frac{k}{4\pi}\gmu \td A\Big]
\\
\delta_{\gmu}\theta &=& \fsLie{\gmu}\theta
\;=\; -\frac{k}{4\pi}\td\gmu\wedge\delta A
\eeqn
(we have assumed that $\gmu$ is field-independent in the second result). Similarly, for diffeomorphisms, we have $\delta_{\un\xi}A={\cal L}_{\un\xi}A$ and
\beqn
\delta_{\un\xi}\Lagr&=&\fsLie{\un\xi}\Lagr\label{diffactionLagr}
\oseq \td\Big[\frac{k}{4\pi}(i_{\un\xi}A) \td A\Big]
\\
\delta_{\un\xi}\theta &=& \fsLie{\un\xi}\theta
\oseq {\cal L}_{\un\xi}\theta
\eeqn
Comparing these results, it is often said that a diffeomorphism can be regarded as a ``field-dependent gauge transformation". We will not use this interpretation here. 
One often also says that the ``anomaly" of $\theta$ vanishes, where $\Delta_{\un\xi}\theta:=\fsLie{\un\xi}\theta-{\cal L}_{\un\xi}\theta$. So we see that the Lagrangian is neither gauge- nor diff-invariant, nor is the presymplectic 1-form. 

The diffeomorphism Noether charge current is given by 
\beq\label{Diffeo Noether Charge Current}
\algcurrdens{\un\xi}=\td\algchgdens{\un\xi}\oseq\fscon{\un\xi}\theta-i_{\un\xi}\Lagr
\eeq
where the second term is present because of the non-invariance of the Lagrangian; it is required by the stipulation that the Noether current be a total derivative, this being a diff-invariant theory.
This evaluates to 
\beq
\td\algchgdens{\un\xi}&=& \frac{k}{4\pi}tr\Big[ \td\Big((i_{\un\xi}A)A\Big)-2(i_{\un\xi}A) F\Big]
\eeq
Thus we see that on-shell, we can identify the diffeomorphism charge density as
\beq
\algchgdens{\un\xi}=\frac{k}{4\pi}tr\Big((i_{\un\xi}A)A\Big)
\eeq
up to a total derivative. There is a similar modification of the Noether current for gauge transformations: by explicit calculation, we find
\beq
\fscon{\gmu}\theta=\frac{k}{4\pi}tr\Big[\td(\gmu A)-2\gmu F +\gmu\td A\Big]
\eeq 
which is not, even on-shell, a total derivative. This fact can be traced to the non-invariance of the Lagrangian. 
A suitable modification of the current is then
\beq\label{gaugeNoethercurrent}
\algcurrdens{\gmu}=\fscon{\gmu}\theta+\frac{k}{4\pi}tr(\td\gmu\wedge A)
\eeq
and one  reads off the Noether charge density
\beq\label{gaugeChargeDens}
\algchgdens{\gmu}=\frac{k}{4\pi}tr\Big(\gmu A\Big)
\eeq
up to a total derivative. We will later explain the geometric origin of the modification \eqref{gaugeNoethercurrent}. It should be clear that this modification is a special property of Chern-Simons theory, as no such modification is apparently necessary, for example, in 
Yang-Mills theory. Again, it is a consequence of the non-invariance of the Lagrangian.

The rosiness of this picture stops here, as it is not the case that the symmetries are represented by Hamiltonian vector fields on field space -- there is an issue with integrability. A Hamiltonian vector field satisfies
\beq
\fscon{\un\xi}\Omega_\Sigma=-\delta\algchg{\un\xi}
\eeq
for some function $\algchg{\un\xi}$ on the phase space of fields, and
where $\Omega_\Sigma$ is the symplectic structure
\beq
\Omega_\Sigma:=\int_\Sigma \phi_1^*(\delta\theta)
\eeq
for an embedded hypersurface $\phi_1:\Sigma\to M$. One finds however that 
\beq
\fscon{\un\xi}\Omega_\Sigma \oseq
\int_\Sigma\phi_1^*\Big(\td i_{\un\xi}\theta-\delta \td\algchgdens{\un\xi}\Big)
\eeq
We would like to read off that $\algchg{\un\xi}=\int_\Sigma\phi_1^*(\td \algchgdens{\un\xi})=\int_{\pa\Sigma}\phi_2^*(\algchgdens{\un\xi})$ but the problem of course is that $i_{\un\xi}\theta$ is neither zero nor a total variation. In \cite{Ciambelli:2021nmv}, we resolved this problem for all diff-invariant theories by extending the phase space by recognizing that the embedding map $\phi_1$ is not in general invariant under diffeomorphisms. The simplest way to understand this 
is to define
\beq \label{Sec 2 Extended Potential}
\theta^{ext.}=\theta+i_{\un\chi}\Lagr
\eeq
where $\un\chi$ is a vector-valued $(0,1)$-form satisfying $\fscon{\un\xi}\un\chi=-\un\xi$ and $\delta\un\chi=-\frac12[\un\chi,\un\chi]$. 
Defining
\beq
\Omega^{ext.}_\Sigma=\delta\Theta^{ext.}_\Sigma=\delta\int_\Sigma\phi_1^*(\theta^{ext.})
\eeq
we find that 
\beq
\fscon{\un\xi}\Omega^{ext.}_\Sigma \oseq-\delta\int_\Sigma\phi_1^*\Big( \td\algchgdens{\un\xi}\Big)
\eeq
so that
\beq
\algchg{\un\xi}=\int_{\pa\Sigma}\phi_2^*(\algchgdens{\un\xi})
\eeq
where $\phi_2:\pa\Sigma\to M$. Furthermore, one then finds that 
\beq
\fscon{\un\xi}\fscon{\un\eta}\Omega^{ext.}_\Sigma=\fscon{[\un\xi,\un\eta]} \int_\Sigma\phi_1^*(\theta^{ext.})
\eeq
and so one has a representation of the diffeomorphisms via Poisson brackets {\em without central extension}.
More precisely, only a subalgebra of the diffeomorphism algebra supports non-zero charges, which is called the {\it extended corner symmetry} \cite{Ciambelli:2021vnn,Ciambelli:2022cfr}.
We should emphasize here that one of the central reasons that extending the phase space in the fashion described here is so important is that it prepares for a quantum theory: any degrees of freedom required for gluing of the physics on arbitrary subregions, for example, are now present. Thus we are capable of properly describing important properties of a quantum theory, such as entanglement and unitarity. 

The reader may find this result puzzling, as it is common knowledge that Chern-Simons theory in 3d gives rise to a centrally extended Virasoro algebra on a boundary.  The above classical result is not in fact inconsistent with the emergence of a central extension of the quantum algebra. The existence of a central extension is a topological issue on the space of fields, and what we have done here by extending the phase space of the theory is to saturate the classical anomaly. In a forthcoming publication \cite{Klinger:2023abc}, we will explain how to construct this properly in a fashion amenable to a quantum theory, building on the configuration algebroid introduced later in the present paper.

Now let us consider the integrability of the gauge charges. So we compute
\beqn
\fscon{\gmu}\Omega_\Sigma&=&\int_\Sigma\phi_1^*\Big( \fscon{\gmu}\delta\theta\Big)
\\
&=&\frac{k}{4\pi}\int_\Sigma\phi_1^*\Big(tr(-2\tD\gmu\wedge \delta A)\Big)
\\
&=&\int_\Sigma\phi_1^*\Big(\frac{k}{2\pi}tr(\mu\delta F)-\delta \td(\algchgdens{\gmu}+\frac{k}{4\pi}tr(\mu A))\Big)
\eeqn
It appears then that if $\delta F=0$, we could redefine the gauge charge density \eqref{gaugeChargeDens} (effectively multiplying it by 2) in order to achieve integrability. This gives 
\beq
\algchgdens{\mu}'=\frac{k}{2\pi}tr(\mu A),\qquad
\{\algchg{\mu}',\algchg{\nu}'\}:=\delta_\nu\algchg{\mu}'=\algchg{-[\mu,\nu]}'+\frac{k}{2\pi}\int_{\pa\Sigma}tr \mu \td\nu
\eeq
This seems to resolve the problem: the charge is integrable and the algebra closes up to a central term. However, $\algchgdens{\mu}'$ is {\it not} the Noether charge density and does not generate the symmetry. It also does not satisfy the correct algebra with the diffeomorphism charge (we expect $\delta_{\un\xi}\algchg{\mu}'=-\algchg{{\cal L}_{\un\xi}\mu}'$). 

Instead, the proper thing to do is extend the phase space again to account for the gauge-noninvariance. To do so, we introduce $\lambda$, a Lie-algebra valued $(0,1)$-form, that satisfies $\fscon{\gmu}\lambda=-\gmu$ and $\delta\lambda=\frac12[\lambda,\lambda]$ in terms of which we write 
\beq\label{sec 2 extended potential 2}
\theta^{ext.}=\theta+i_{\un\chi}\Lagr+\frac{k}{4\pi}\td tr(\lambda\td A)
\eeq
where in addition we have $\fscon{\gmu}\un\chi=0=\fscon{\un\xi}\lambda$ and $\fscon{\un\xi}\un\chi=-\un\xi$ as before.
We compute then that
\beqn\label{Gauge Current}
\fscon{\mu}\theta^{ext.}&=&\fscon{\mu}\theta-\frac{k}{4\pi}tr(\mu \td A)
=\frac{k}{4\pi} \td\Big(tr(\mu A)\Big)
-\frac{k}{2\pi} tr(\mu F)
\;\;\hat{=}\;\;\td\algchgdens\mu
\\\label{Diff Current}
\fscon{{\un\xi}}\theta^{ext.}
&=&\fscon{{\un\xi}}\theta-i_{\un\xi}L
=\frac{k}{4\pi}\td\Big(tr(i_{\un\xi}A) A\Big)
-\frac{k}{2\pi}tr(i_{\un\xi}A) F
\;\;\hat{=}\;\;\td\algchgdens{\un\xi}
\eeqn
We will write $\Theta^{ext.}_\Sigma=\int_\Sigma \phi_1^*(\theta^{ext.})$. Given that $\fsLie{\un\xi}\Theta^{ext.}_\Sigma=\int_\Sigma \phi_1^*(\Delta_{\un\xi}\theta^{ext.})=0=\fsLie{\gmu}\Theta^{ext.}_\Sigma=\int_\Sigma \phi_1^*(\Delta_{\gmu}\theta^{ext.})$, the Noether charges are thus integrable, \beq
\fscon{\mu}\Theta^{ext.}_\Sigma \oseq \algchg{\mu},\qquad
\fscon{{\un\xi}}\Theta^{ext.}_\Sigma\oseq \algchg{\un\xi}
\eeq
and by direct computation we eventually find
\beqn \label{Charge Algebra}
\fscon{\gnu}\fscon{\gmu}\Omega^{ext.}_\Sigma \oseq \algchg{[\gmu,\gnu]},\qquad
\fscon{\un\xi}\fscon{\gmu}\Omega^{ext.}_\Sigma=-\fscon{\gmu}\fscon{\un\xi}\Omega^{ext.}_\Sigma\oseq \algchg{{\cal L}_{\un\xi}\gmu}
,\qquad
\fscon{\un\eta}\fscon{\un\xi}\Omega^{ext.}_\Sigma \oseq\algchg{[\un\eta,\un\xi]}
\eeqn
We thus see that the extension of the phase space yields not only integrable charges, but a representation of the $\text{Diff}(M)\ltimes G$ algebra. Without extending phase space, one apparently cannot attain such a representation, centrally extended or not. In the next sections, we will show that the above extensions of the phase space fit neatly together within the Lie algebroid structure underlying the Chern-Simons gauge theory. In fact, the algebroid construction so formulated will be applicable to {\it any} theory with local symmetries.

\section{Atiyah Lie algebroids}\label{sec:Primer}

In this section we will provide an overview of the prerequisite knowledge about Lie algebroids which will be utilized in the main argument of the paper. For a more complete introduction to these ideas see \cite{Ciambelli:2021ujl,Jia:2023tki}. 

We begin with a principal bundle $P(M,G)$, where $M$ is the base manifold, and $G$ is the structure group. We denote by $\pi: P \rightarrow M$, and $R: G \times P \rightarrow P$ the projection and free right action, respectively. These are the {canonical maps} which define $P(M,G)$. The \emph{Atiyah Lie algebroid derived from the principal bundle $P(M,G)$} is given by the vector bundle $A \equiv P \times_G TP=TP/G \rightarrow M$, which inherits its bracket $[\cdot,\cdot]_A$ and anchor $\rho = \pi_*: A \rightarrow TM$ from $TP$. The map $\rho$ is automatically surjective, and therefore the algebroid $A$ is \emph{transitive}. This means that there exists a short exact sequence of vector bundles over $M$:
\begin{equation} \label{Short Exact Sequence 1}
\begin{tikzcd}
0
\arrow{r} 
& 
L
\arrow{r}{j} 
&
A
\arrow{r}{\rho} 
& 
TM
\arrow{r} 
&
0 \,.
\end{tikzcd}
\end{equation}
Here, $L = P \times_G \mathfrak{g} \rightarrow M$ is the kernel of the map $\rho$, and can be seen to be a bundle of Lie algebras with standard fiber $\mathfrak{g}$. The pair of maps $(\rho,j)$ are the {canonical maps} of the Atiyah Lie algebroid $A$, and can be regarded as descending from $(\pi,R)$. 

The  maps $(\rho,j)$ identify the {vertical sub-bundle} of $A$ as $V = \text{im}(j) = \text{ker}(\rho)$. An Atiyah Lie algebroid subsequently possesses a {horizontal distribution} $H$ such that $A = H \oplus V$. The definition of such a horizontal distribution is \emph{not} canonical, and is associated with the choice of a {connection} on $A$. A connection on $A$ is a second short exact sequence:
\begin{equation} \label{Short Exact Sequence 2}
\begin{tikzcd}
0
\arrow{r} 
& 
L
\arrow{r}{j} 
\arrow[bend left]{l} 
& 
A
\arrow{r}{\rho} 
\arrow[bend left]{l}{\omega}
& 
TM
\arrow{r} 
\arrow[bend left]{l}{\sigma}
&
0\,,
\arrow[bend left]{l} 
\end{tikzcd}
\end{equation}
such that $H = \text{im}(\sigma) = \text{ker}(\omega)$. 

A connection on $A$ induces a {Lie algebroid representation} on vector bundles $E \rightarrow M$ possessing a linear representation of the structure group. In general, a Lie algebroid representation of $E$, $\Aconn{E}: A \rightarrow \text{Der}(E)$ is a {morphism}\footnote{The map $\Aconn{E}: A \rightarrow \text{Der}(E)$ is a morphism if:
\beq
R^{\Aconn{E}}(\mX,\mY) = [\Aconn{E}(\mX),\Aconn{E}(\mY)]_{\text{Der(E)}} - \Aconn{E}([\mX,\mY]_A) = 0, \;\;\; \forall \; \mX,\mY \in A.
\eeq
The bracket on $\text{Der}(E)$ is defined with respect to the usual composition of operators. Given $A, B \in \text{Der}(E)$, or equivalently, $A,B: \Gamma(E) \rightarrow \Gamma(E)$, we have, for all $\un{\psi} \in \Gamma(E)$, $AB(\un{\psi}) = A \circ B(\un{\psi})$. Thus, $[A,B]_{\text{Der}(E)}(\un{\psi}) = A \circ B(\un{\psi}) - B \circ A(\un{\psi})$.} compatible with the anchor $\rho$\footnote{Compatibility with the anchor simply implies that $\Aconn{E}$ is a derivation in the Leibniz sense:
\beq
\Aconn{E}(\mX)(f\un{\psi}) = f\Aconn{E}(\mX)(\un{\psi}) + \rho(\mX)(f)\un{\psi}, \;\;\; \forall \; \mX \in A, \; f \in C^{\infty}(M), \; \un{\psi} \in E.
\eeq}, which can be regarded as specifying a notion of differentiating sections of $E$ along sections of $A$. The Lie algebroid representation induced by the connection $(\sigma,\omega)$ is given by
\beq \label{Lie algebroid rep}
	\Aconn{E}(\mX)(\un{\psi}) = \nabla^E_{\rho(\mX)} \un{\psi} - v_E \circ \omega(\mX)\un{\psi}.
\eeq
Here, $\nabla^E: TM \rightarrow \text{Der}(E)$ is a covariant derivative, and $v_E: L \rightarrow \text{End}(E)$ is an endomorphism. The fact that $\Aconn{E}$ is a Lie algebroid representation implies that $v_E$ is a linear representation, and that the curvature of $\nabla^E$ is determined entirely by the curvature of $H \subset A$ through the equation
\beq
	R^{\nabla^E}(\un{X},\un{Y}) = [\nabla^E_{\un{X}}, \nabla^E_{\un{Y}}]_{\text{Der}(E)} - \nabla^E_{[\un{X},\un{Y}]_{TM}} = -v_E \circ \omega(R^{\sigma}(\un{X},\un{Y})).
\eeq
The right-hand side can be thought of as a matrix representation of the curvature.

Let $\Omega(A;E) = \bigoplus_{p = 1}^{\text{rank}(A)} \Omega^p(A;E)$ denote the exterior algebra of $A$ with values in $E$. Here $\Omega^p(A;E) \equiv \wedge^p A^* \otimes E$ consists of all totally antisymmetric $p$-linear maps from $A$ into $E$. The Lie algebroid representation $\Aconn{E}$ induces a coboundary operator $\hatd: \Omega^p(A;E) \rightarrow \Omega^{p+1}(A;E)$ defined by the {Koszul formula}. Explicitly, given $\eta \in \Omega^p(A;E)$ and $\{\mX_i\}_{i = 1}^{p+1} \in A$:
\begin{align} \label{dhat on E}
    \hatd \eta(\mX_1, ..., \mX_{p+1}) ={}& \sum_{i} (-1)^{i+1} \Aconn{E}(\mX_i) \eta(\mX_1, ..., \widehat{\mX}_i, ..., \mX_{p+1}) \\
    &+ \sum_{i < j} (-1)^{i + j} \eta([\mX_i, \mX_j]_A, \mX_1, ..., \widehat{\mX}_i, ..., \widehat{\mX}_j, ..., \mX_{p+1}) \,.
\end{align}
The operator $\hatd$ is automatically {nilpotent} due to the fact that $\Aconn{E}$ is a morphism, and $[\cdot,\cdot]_A$ satisfies the {Jacobi identity}. Given the operator $\hatd$ we can now define the {curvature reform}:\footnote{We have introduced the graded Lie bracket between $L$-valued differential forms. For $\alpha \in \Omega^m(A; L)$ and $\beta \in \Omega^n(A; L)$, $[\alpha\wedge\beta]_{L}$ is defined as
\begin{equation}
	[\alpha\wedge\beta]_{L}(\mX_1, \cdots, \mX_{m+n}) = \sum_{\sigma }\text{sgn}(\sigma) [\alpha(\mX_{\sigma(1)}, \cdots, \mX_{\sigma(m)}), \beta(\mX_{\sigma(m+1)}, \cdots, \mX_{\sigma(m+n)})]_{L}\,,
\end{equation}
where $\mX_1\cdots \mX_{m+n}$ are arbitrary sections on $A$, $\sigma$ denotes the permutations of $(1,\cdots,m+n)$, and $\text{sgn}(\sigma)=1$ for even permutations and  $\text{sgn}(\sigma)=-1$ for odd permutations.}
\beq
	\Omega = \hatd \omega + \frac{1}{2}[\omega \wedge \omega]_L \in \Omega^2(A;L),
\eeq
which can be shown to be fully {horizontal}. It is also related to the curvature of the horizontal distribution through the series of equations
\beq
	\Omega(\mX,\mY) = -R^{-\omega}(\sigma\circ \rho(\mX),\sigma\circ\rho(\mY)) = -\omega(R^{\sigma}(\rho(\mX),\rho(\mY))).
\eeq

Two Lie algebroids $A_1$ and $A_2$ are said to be {morphic} if there exists a \hlt{Lie algebroid morphism}\footnote{Generically speaking, a Lie algebroid morphism is a map between two algebroids which preserves brackets. Hereafter, although we use the word morphism, we consider maps $\varphi$ which satisfy the stronger condition of being an isomorphism between the vector bundles on which the algebroids are based in addition to preserving brackets.} $\varphi: A_1 \rightarrow A_2$ along with an induced isomorphism $g_{\varphi}: E_1 \rightarrow E_2$ intertwining associated Lie algebroid representations. Let $\fldspalg$ denote the set of of all Atiyah Lie algebroids with connection derived from principal bundles with base manifold $M$ and structure group $G$, along with their associated vector bundles and exterior algebras. For any two such algebroids, $A_1$ and $A_2$ there exists a map $\varphi: A_1 \rightarrow A_2$ defined by the commutative diagram
\begin{equation}
\label{Transitive Lie Algebroid Morphism}
\begin{tikzcd}
& 
L_1
\arrow[bend left]{r}{j_1}
\arrow{dd}{v}
& 
A_1
\arrow[left]{dd}{\varphi}
\arrow[left]{l}{\omega_1}
\arrow[bend left]{r}{\rho_1}
& 
TM_1
\arrow[shift left]{l}{\sigma_1} 
\arrow{dd}{\phi_*}
&
\\
0
\arrow{dr}
\arrow{ur}
&
&
&
&
0\,.
\arrow{ul} 
\arrow{dl} 
\\
& 
L_2
\arrow[bend right, swap]{r}{ j_2}
\arrow[shift left]{uu}{\bar v}
&
A_2
\arrow[left, swap]{l}{\omega_2}
\arrow[bend right, swap]{r}{ \rho_2}
\arrow[shift left]{uu}{\overline{\varphi}}
&
TM_2
\arrow[swap]{l}{\sigma_2} 
\arrow[shift left]{uu}{\bar\phi_*}
&
\end{tikzcd}
\end{equation}
Here, we have included a diffeomorphism $\phi:M_1\to M_2$ and a morphism $v:L_1\to L_2$. 
In \cite{Jia:2023tki}, we presented a simplified version of this commutative diagram, where we regarded $M_1\sim M_2\sim M$ and $L_1\sim L_2\sim L$.

In \cite{Jia:2023tki}, it is shown there that $\varphi$ defined by \eqref{Transitive Lie Algebroid Morphism} has curvature.  
\beq
	R^{\varphi}(\mX,\mY) = j_2\circ v\left((\varphi^*\Omega_2)(\mX,\mY) - \Omega_1(\mX,\mY)\right).
\eeq
Here we have made use of the Lie algebroid pullback, $\varphi^*: \Omega(A_2;E_2) \rightarrow \Omega(A_1;E_1)$. Explicitly, given $\eta \in \Omega^r(A_2;E_2)$ and $\mX_1, ..., \mX_r \in A_1$ we have
\begin{equation}\label{Lie algebroid pullback}
	(\varphi^*\eta)(\mX_1, ..., \mX_r) =g_\varphi^{-1}\Big(\eta(\varphi(\mX_1), ..., \varphi(\mX_r))\Big)\,. 
\end{equation}
Thus, a necessary and sufficient condition for \eqref{Transitive Lie Algebroid Morphism} to define a Lie algebroid morphism is for the curvatures of the horizontal distributions of $A_1$ and $A_2$ to be ``equal" in the sense that $\Omega_1 = \varphi^*\Omega_2$. This observation suggests that the set of Lie algebroid morphisms, $\morphalg$, should be identified with the set of diffeomorphisms on $M$ and gauge transformations in $L$. This suggestion is brought to fruition in \cite{Jia:2023tki} where it is shown that the action of $\morphalg$ on $\fldspalg$ reproduces the familiar gauge transformation formulae. 

Recall that in \cite{Ciambelli:2021nmv}, the extended phase space in the context of diffeomorphism-invariant theories was associated with the physicality of an embedding map. This can be thought of in terms of certain restricted diffeomorphisms. Given that $TM$ can be thought of as a Lie algebroid with trivial isotropy, what we are describing here is a natural generalization. In fact, we are proposing that the appropriate generalization of the embedding map is a Lie algebroid morphism. We will show in the following sections that this works properly; indeed one can think of a morphism as consisting locally of a diffeomorphism and a gauge transformation.\footnote{Note that one of the results of the commutativity of \eqref{Transitive Lie Algebroid Morphism} is $\rho_2\circ\varphi=\phi_*\circ\rho_1$. This implies that $\varphi$ carries additional freedom beyond the diffeomorphism $\phi$; essentially any gauge transformation will do.} In the context of defining a Noether charge, this freedom can be thought of as corresponding to a choice of embedding along with a choice of gauge. See Appendix C for a detailed discussion.

It can moreover be shown that if $\varphi$ is a morphism, it will also be a \emph{chain map} in the exterior algebra sense. That is
\beq \label{Chain Map condition}
	\hatd_1 \circ \varphi^* = \varphi^* \circ \hatd_2.
\eeq
This fact along with the fact that the curvatures of $A_1$ and $A_2$ are equivalent up to a pullback imply that morphic algebroids are topologically equivalent in the sense that they have isomorphic cohomology classes. 

The culmination of these observations is to define a relation under Lie algebroid morphism such that $(A_1,\omega_1) \sim (A_2,\omega_2)$ provided there exists a Lie algebroid morphism $\varphi: A_1 \rightarrow A_2$. This relation is naturally an equivalence relation since it is clearly \emph{reflexive}, it is \emph{symmetric} due to the fact that every morphism $\varphi: A_1 \rightarrow A_2$ gives rise to an inverse morphism $\overline{\varphi}: A_2 \rightarrow A_1$, and it is \emph{transitive} since the composition of two morphisms is itself a morphism.\footnote{This is proven, for example, by observing that the curvature of the composition of two maps $\varphi_{12}: A_1 \rightarrow A_2$ and $\varphi_{23}: A_2 \rightarrow A_3$ is given by 
\beq
	R^{\varphi_{23} \circ \varphi_{12}}(\mX,\mY) = R^{\varphi_{23}}(\varphi_{12}(\mX),\varphi_{12}(\mY)) + \varphi_{23}(R^{\varphi_{12}}(\mX,\mY)).
\eeq
Hence $R^{\varphi_{23} \circ \varphi_{12}} = 0$ if $R^{\varphi_{23}} = R^{\varphi_{12}} = 0$.
} Let
\beq
	[(A,\omega)] \equiv \{(A',\omega') \in \fldspalg \; | \; (A',\omega') \sim (A,\omega)\}
\eeq
denote the equivalence class of $(A,\omega) \in \fldspalg$. The set of all Atiyah Lie algebroids derived from principal bundles with base manifold $M$ and structure group $G$ is therefore split into a collection of such equivalence classes. 

In \cite{Jia:2023tki} it was shown that every equivalence class $[(A,\omega)]$ possesses a natural \emph{local} representative in terms of the \emph{trivialized Lie algebroid}, $(A_{\tau},\omega_{\tau})$. In any open set $U \subset M$, the trivialized Lie algebroid is of the form $A_{\tau} = TU \oplus L^{U}$. For each $(A,\omega)$, and any such open set $U \subset M$, there exists a morphism $\tau: A^U \rightarrow A_{\tau}$ which we refer to as a \emph{Lie algebroid trivialization}.\footnote{For simplicity we will not distinguish the open set $U$, with the understanding that the following is valid in any such open neighborhood. We use the subscript $\tau$ hereafter to remind the reader when a given statement is valid in a local trivialization.} Despite its appearance the trivialized Lie algebroid \emph{does not} have a trivial connection. Indeed, for $\tau$ to be a morphism it must be true that $A_{\tau}$ possesses a horizontal distribution with curvature $\Omega_{\tau}$ such that for all $(A,\omega) \in [(A_\tau,\omega_\tau)]$ with curvature $\Omega$, $\Omega = \tau^* \Omega_{\tau}$. The connection reform of the trivialized algebroid can be explicitly related to a gauge field $b \in \Omega^1(M;L)$ and a ghost field\footnote{Note that here we refer to $\varpi$ as the ghost field. However, we are not using the common physics language based on Grassmann algebra --- in the Atiyah Lie algebroid construction, $\varpi$ is instead a field of Maurer-Cartan forms. See \cite{Jia:2023tki} for more detailed discussion of this and the corresponding BRST structure.} $\varpi \in \Omega^1(L;L)$ such that \cite{Jia:2023tki} 
\beq
	\omega_{\tau} = b - \varpi.
\eeq
The curvature $\Omega_{\tau}$ is therefore related to the gauge field strength of $b$ and will not be zero unless $\Omega = 0$ for all $(A,\omega) \in [(A_\tau,\omega_\tau)]$. 

Given the form of the trivialized Lie algebroid, it is natural to formulate the exterior algebra of an algebroid $(A,\omega) \in [(A_{\tau},\omega_{\tau}]$ as a bicomplex $\Omega(A_{\tau};E) = \Omega(M,L;E)$. Here
\beq \label{Consistent splitting}
	\Omega^p(M,L;E) = \bigoplus_{r + q = p} \Omega^{(r,q)}(M,L;E),
\eeq
where $\Omega^{(r,q)}(M,L;E)$ consists of totally anti-symmetric $p$-linear maps from $r$ copies of $TM$ and $q$ copies of $L$ into $E$. In other words, $\Omega(M,L;E)$ is generated by a basis of one forms in $T^*M$, $\{dx^{\mu}\}_{\mu = 1}^{\text{dim}(M)}$, a basis of dual elements in $L^*$, $\{t^A\}_{A = 1}^{\text{dim}(G)}$, and a basis of sections for $E$, $\{\un{e}_a\}_{a = 1}^{\text{rank}(E)}$. We will often refer to \eqref{Consistent splitting} as the consistent splitting of the exterior algebra. By contrast, the formulae involving \eqref{Lie algebroid rep} may be regarded as existing in a covariant splitting of the algebroid because such equations respect the covariant horizontal-vertical splitting of $A$. 

Since cohomology is preserved under Lie algebroid morphism, one can analyze the exterior algebra of $(A,\omega) \in [(A_{\tau},\omega_{\tau})]$ by using $(A_{\tau},\omega_{\tau})$ as a representative. In \cite{Jia:2023tki} it was shown that the action of $\hatd_{\tau}$ takes a relatively simple and physically transparent form, which makes the requisite computations particularly straightforward. We recount these results now, and refer the reader to \cite{Jia:2023tki} for a more in depth calculation. 

Here we will recall from \cite{Jia:2023tki} the action of $\hatd_\tau$ on a selection of elements of $\Omega(M,L;E)$.  Acting on a zero-form in a representation space, $\un{\psi} \in E$, we may write
\beq \label{dhat tau on zero form}
	\hatd_{\tau} \un{\psi} = \td \psi^a \otimes \un{e}_a + v_E(\varpi)(\un{\psi}).
\eeq
Because $A_\tau$ possesses a connection, the first term gives rise to a covariant derivative: this can be displayed by contracting with a horizontal section 
\beq
	\hatcon{\sigma_{\tau}(\un X)} \hatd_{\tau} \un{\psi} =  \nabla^E_{\un X}\un{\psi} = \delta_{\un X} \un{\psi}=\hatLie{\sigma_{\tau}(\un X)}\un{\psi},\qquad \un X\in\Gamma(TM)
\eeq
We note that while this is the Lie derivative on the algebroid $A_\tau$, it is the gauge covariant derivative from the perspective of $TM$.
The meaning of the second term in \eqref{dhat tau on zero form} is best elucidated by contracting with a vertical element in $A_{\tau}$. In particular,
\beq
	\hatcon{-j_{\tau}(\gmu)} \hatd_{\tau} \un{\psi} =  v_E(-\gmu)\un{\psi} = \delta_{\gmu} \un{\psi}=\hatLie{-j_{\tau}(\gmu)}\un{\psi},\qquad \gmu\in\Gamma(L).
\eeq
That is, the vertical piece of $\hatd_{\tau}$ encodes the gauge transformation of $\un{\psi}$. The action of $\hatd_{\tau}$ on the gauge field $b \in \Omega^1(M;L)$ and the Maurer-Cartan form $\varpi \in \Omega^1(L;L)$ are given by
\beq \label{hatd tau on b and varpi}
	\hatd_{\tau} b = \td b + \td \varpi^A \otimes \un{t}_A + [\varpi \wedge b]_L, \qquad \hatd_{\tau} \varpi = \td \varpi^A \otimes \un{t}_A + \frac{1}{2}[\varpi \wedge \varpi]_L.
\eeq
Again, \eqref{hatd tau on b and varpi} can benefit from some unpacking, particularly the term $\td\varpi^A \otimes \un{t}_A$. Contracting with $-j_{\tau}(\gmu) \in A_{\tau}$ we find
\beq \label{de Rham of varpi}
	\hatcon{-j_{\tau}(\gmu)} (\td \varpi^A \otimes \un{t}_A) = \td \mu^A \otimes \un{t}_A. 
\eeq
This implies that
\beq
	\hatcon{-j_{\tau}(\gmu)} \hatd_{\tau} b = \td \mu^A \otimes \un{t}_A + [b,\gmu]_L = \delta_{\gmu} b=\hatLie{-j_{\tau}(\gmu)} b,
\eeq
that is, the vertical part of $\hatd_{\tau}$ encodes the gauge transformation of $b$. 

An important corollary of \eqref{de Rham of varpi} is that it ensures the Lie derivative of the Maurer-Cartan form along $j_{\tau}(\gmu) \in A_{\tau}$ is zero at any point in $M$. This was to be expected by the definition of $\varpi$ as an invariant form. To be explicit, we can compute:
\beq \label{Lie of MC = 0}
\delta_{\gmu}\varpi=	\hatLie{-j_{\tau}(\gmu)} \varpi = \hatd_{\tau} \hatcon{-j_{\tau}(\gmu)} \varpi + \hatcon{-j_{\tau}(\gmu)} \hatd_{\tau} \varpi = -\hatd_{\tau} \gmu + \left(d\mu^A \otimes \un{t}_A + [\varpi,\gmu]_L\right) = 0.
\eeq
To arrive at the conclusion of \eqref{Lie of MC = 0} we have used \eqref{dhat tau on zero form} applied to a section of $L$, and interpreted $[\varpi,\gmu]_L = v_L(\varpi)(\gmu)$, where the linear representation is given by the adjoint action of $L$ on itself \cite{Ciambelli:2021ujl,mackenzie2005general}. 

\section{The Configuration Algebroid}
\label{sec:confalg}

In the preceding section we introduced $\fldspalg$ as the set of all Atiyah Lie algebroids derived from principal bundles sharing the same base manifold and structure group, along with sections of their associated vector bundles and exterior algebras. We also saw that there is a natural action on $\fldspalg$ by $\morphalg$, the set of all Lie algebroid morphisms, and, moreover, that this action reproduces the action of local diffeomorphisms and gauge transformations on exterior algebra sections. An element of $\fldspalg$ is given by $(A,\Phi)$, where $A$ is an Atiyah Lie algebroid, and $\Phi$ is symbolic notation for a particular configuration in the set of all sections of $\Omega(A;E)$, including the connection reform $\omega$.\footnote{As we shall see, this picture implies that we are treating all of the possible degrees of freedom of $A$ as dynamical. However, we can relegate particular degrees of freedom to the background by withholding them from our specification of $\fldspalg$. This would be relevant, for example, if we wanted to consider a field theory on a particular metric background.} $\fldspalg$ may be regarded as the algebroid promotion of the usual notion of the space of all field configurations \cite{Crnkovic:1986ex,Crnkovic:1987tz,Compere:2018aar,cmp/1103899589,GAWEDZKI1972307,cmp/1103858807}. For brevity, we will drop the specification of $A$ in $(A,\Phi)$, and refer to elements of $\fldspalg$ simply by the field configuration $\Phi$. As we have introduced in Section \ref{sec:Primer}, the differentiation between equivalence classes of morphic algebroids is characterized by whether or not their connections are gauge equivalent. Hence, we can regard the identification of different algebroids as given by the specification of the connection reform which is included in $\Phi$. 

In traditional accounts of gauge theories \cite{Faddeev:1967fc,gribov1978quantization}, one would then attempt to quotient $\fldspalg$ by $\morphalg$ in order to obtain the space of gauge orbits, which we will call $\spalg_0$. Equivalently, $\fldspalg$ is thought of as a principal $\morphalg$-bundle over $\spalg_0$ \cite{deAGomes:2008grp,Prabhu:2015vua,Gomes:2016mwl}. In this paper, our intention instead is to describe how to implement the extended configuration space, as we have seen is necessary in general in Section 2. 
We will demonstrate that this extension can be formulated by first constructing a principal $\morphalg$-bundle over $\fldspalg$. To this end, we define the \hlt{extended configuration space}  to be the set (locally)
\beq \label{Extended configuration space}
	\pbalg := \{(\varphi,\Phi) \; | \; \varphi \in \morphalg, \; \Phi \in \fldspalg\},
\eeq
which we will  show is a principal $\morphalg$-structure. The projection of $\pbalg$ is defined by
\beq
	\pi: \pbalg \rightarrow \fldspalg; \qquad (\varphi, \Phi) \mapsto \varphi^* \Phi.
\eeq
We define the right action of $\morphalg$ on $\pbalg$ by\footnote{We should recognize that $\morphalg$ is not a group but rather a groupoid. However, just as a Lie group has an associated Lie algebra, a Lie groupoid has an associated Lie algebroid. In our context this Lie algebroid can be identified with $A$ itself, a fact that we will make use of in what follows.}
\beq
	R: \morphalg \times \pbalg \rightarrow \pbalg; \qquad (\varphi',(\varphi,\Phi)) \mapsto (\varphi' \circ \varphi, (\overline{\varphi}')^* \Phi).
\eeq
Notice that the projection is invariant under the right action:
\beq
    \pi \circ R_{\varphi'}((\varphi,\Phi)) = \pi(\varphi' \circ \varphi, (\overline{\varphi}')^*\Phi) = \pi(\varphi, \Phi),
\eeq
as is necessary for a well defined $\morphalg$-structure. This demonstrates that $\pbalg(\fldspalg,\morphalg)$ is indeed a principal $\morphalg$-structure. 

The reader may be puzzled, since we are advocating that it is productive to, in a sense, {\it multiply} by $\morphalg$ rather than divide. As an analogy, we might consider an ordinary manifold $M$; diffeomorphisms can be thought to act on points of $M$, generating orbits, much like morphisms act on points in $\fldspalg$ (field configurations) generating orbits. Here, the analogue of constructing $\pbalg$ is to construct $TM$, whereby diffeomorphisms are thought to be generated by sections. This isn't quite right, because $TM$ is not a principal bundle; however, it can be thought of as an algebroid, and in fact our next step will be to define the Atiyah Lie algebroid corresponding to $\pbalg$.

\disc{
}



Following this path, we now carry over the analysis of Section \ref{sec:Primer} to define the Atiyah Lie algebroid derived from $\pbalg(\fldspalg,\morphalg)$: 
\beq \label{Short Exact Sequence for Configuration Algebroid}
\begin{tikzcd}
0
\arrow{r} 
& 
\ellalg = \pbalg \times_{\text{Ad}} A
\arrow{r}{\alginjo} 
& 
\algalg = \Tpbalg/\morphalg
\arrow{r}{\algrho} 
& 
\text{T}\fldspalg
\arrow{r} 
&
0
\end{tikzcd}
\eeq
We will refer to $\algalg$ as the \hlt{configuration algebroid}. 
An especially important observation made in \eqref{Short Exact Sequence for Configuration Algebroid} is that the isotropy bundle of $\algalg$ is a bundle of Lie algebroids over $\fldspalg$. In other words, at each $\Phi \in \fldspalg$, $\ellsec{\mu} \in \ellalg$ takes values in $A$ and depends on the fields $\Phi$. More to the point, since $\ellalg$ is the local algebroid integrating to $\morphalg$, we can regard its elements as generating infinitesimal, (generally field-dependent) diffeomorphisms and gauge transformations. That is, $\ellsec{\mu} \in \ellalg$ is determined by a section of $A$ at a point $\Phi\in\fldspalg$, which can be regarded as a pair $(\un\xi,\gmu)$, where $\un\xi\in\Gamma(TM)$ and $\gmu\in\Gamma(L)$. The anchor space of $\algalg$ is the space of diffeomorphisms on $\fldspalg$. We note in passing that the modified bracket\footnote{Often, field dependence arises upon the imposition of some boundary conditions. Here, field dependence arises generically because of the bundle structure over the space of fields. Hence a field-independent section of $\ellalg$ is given by a (locally) constant section.} often considered in the literature \cite{Chandrasekaran:2020wwn,Barnich:2009se,Barnich:2010eb,Barnich:2010ojg,Barnich:2011mi,Barnich:2013axa,Troessaert:2015nia,Wieland:2021eth},
 appears here as simply the bracket on $\algalg$, given the trivialization of $\algalg$ discussed below. 

It is also perfectly natural for the configuration algebroid to possess a non-trivial connection\footnote{The idea of a connection on the space of fields was advocated for in \cite{Gomes:2016mwl,GOMES2019249} in relation with an associated ``supermetric".}, specified by a short exact sequence
\beq \label{Short Exact Sequence for Configuration Algebroid with Connection}
\begin{tikzcd}
0
\arrow{r} 
& 
\ellalg
\arrow{r}{\alginjo} 
\arrow[bend left]{l}
& 
\algalg
\arrow{r}{\algrho}
\arrow[bend left]{l}{\mathbbb{w}}
& 
\text{T}\fldspalg
\arrow{r}
\arrow[bend left]{l}{\mathbbb{s}} 
&
0
\arrow[bend left]{l}
\end{tikzcd}.
\eeq
In this paper  the role of the connection on $\algalg$ will not play a central role. Rather, in order to make contact with more familiar ideas we will make use of the consistent splitting of the configuration algebroid as introduced in Section \ref{sec:Primer} in the context of a convenient choice of trivialization. So in what follows, when we refer to $\algalg$, we actually mean a local trivialization (which we will not explicitly denote), and our discussion will not directly involve the connection. In a forthcoming publication \cite{Klinger:2023abc} we will elaborate on the important role played by the connection in detail. With that being said, let us again stress that working in the trivialized splitting does not imply that the algebroid we are working with has a trivial connection. 

A generic section of $\algalg$ in its consistent splitting \cite{Jia:2023tki} takes the form
\beq
	\algsec{X} = \algvec{} \oplus \ellsec{\mu},
\eeq
where $\algvec{} \in \text{T}\fldspalg$ and $\ellsec{\mu} \in \ellalg$. When necessary, we take a basis for $\text{T}\fldspalg$ to be given by the set of variational derivatives $\{\frac{\alghatdTM}{\alghatdTM \Phi^I}\}$ where $I$ indexes a set of representations of $A$. 

There is a complication because $\morphalg$ acts non-trivially on $\fldspalg$. This can be fully accounted for by writing $\algrho:\algalg\to T\fldspalg$ \footnote{Note that this wouldn't have happened if we regarded a principal bundle over $\spalg_0$, since the action of $\morphalg$ on $\spalg_0$ is trivial. However, if we had treated $\pbalg$ as a principal bundle over $\spalg_0$, we would not be able to construct the appropriate extension since functionals would have to be written in terms of gauge equivalence classes, rather than arbitrary field configurations.}
\beq
\algrho(\algvec{}\oplus \ellsec{\mu})=\algvec{}+\algvec{\ellsec{\mu}}
\eeq
Here, $\algvec{\ellsec{\mu}}$ is a vector generating the action of $\morphalg$ on $\fldspalg$. 
Furthermore, requiring $\algrho\circ\alginjo=0$, we find
\beq \label{Structure Maps of Configuration Algebroid}
\alginjo(\ellsec{\mu}) =    -\algvec{\ellsec{\mu}}\oplus\ellsec{\mu}.
\eeq

In the course of writing \eqref{Structure Maps of Configuration Algebroid}, we have introduced the vector $\algvec{\ellsec{\mu}}$. Since this generates the action of $A$ on $\fldspalg$, it can be written
\beq \label{V ellsec}
	\algvec{\ellsec{\mu}} = \Aconn{E_I}(\ellsec{\mu}) (\Phi^I) \frac{\alghatdTM}{\alghatdTM \Phi^I} \in \text{T}\fldspalg.
\eeq
At a given point $\Phi \in \fldspalg$, $\Aconn{E_I}$ is the Lie algebroid representation of $A$ acting on the appropriate representation to which $\Phi^I$ belongs. Equation \eqref{V ellsec} therefore represents the generalization\footnote{Specifically, here we mean that $\algvec{\ellsec{\mu}}$ is given, not by the Lie derivative on $\fldspalg$, but by the appropriate representation to which each $\Phi^I$ belongs. In addition, the notation refers to a section of an algebroid $A$ and thus includes both diffeomorphisms of $M$ and gauge transformations. } of $\algvec{\gmu}$ appearing, for example, in eq. \eqref{Gauge transform of A} of Section \ref{sec:conventional}. It quantifies the action of a gauge transformation as carried out by varying the individual field configurations, $\Phi \in \fldspalg$.



The algebroid $\algalg$ possesses an exterior algebra which we  denote by $\Omega(\algalg;\mathbf{E})$. In this paper we will be interested exclusively in $\Omega(\algalg;\Omega(A;E))$, where $A$ is allowed to vary over $\fldspalg$. This is the generalization of the familiar notion of differential forms on field configuration space taking values in the exterior algebra of the underlying spacetime manifold \cite{vinogradov1978spectral,VINOGRADOV19841,VINOGRADOV198441,10.1007/BFb0089725,vinogradov1977algebro,anderson1992introduction,Anderson:1996sc,anderson1996asymptotic}. The exterior algebra $\Omega(\algalg;\Omega(A;E))$ possesses a coboundary operator which we denote by $\alghatd$, along with a  contraction $\algcon{}$ and a Lie derivative $\algLie{}$ such that $\algLie{\algsec{X}} = \alghatd \algcon{\algsec{X}} + \algcon{\algsec{X}} \alghatd$ for any $\algsec{X}\in \Gamma(\algalg)$. Following the analysis of Section \ref{sec:Primer}, we can specify the action of $\alghatd$ in the consistent splitting by generalizing \eqref{dhat tau on zero form}, and \eqref{hatd tau on b and varpi}. Typically we will be interested in functions on the space of fields $\fldspalg$ or in differential forms, which we now interpret as sections of $\Omega(\algalg)$. For our purposes it will be relevant to recall that for any ${\bm \alpha} \in \Omega^{(p,0)}(\fldspalg,\ellalg;\mathbf{E})$, written in a local basis as ${\bm \alpha}=\alpha^a[\Phi] {\bf e}_a$,
\beq\label{deltahatalpha}
	\alghatd {\bm \alpha} = \alghatdTM \alpha^a \otimes {\bf e}_a + \algEnd{\mathbf{E}}(\MCb) ({\bm \alpha}),
\eeq
(a generalization of \eqref{dhat tau on zero form}). 
Here we have introduced the Maurer-Cartan form  $\MCb$ on $\ellalg$, which is defined by
\beq \label{Lie MC = 0}
	\algLie{-\alginjo(\ellsec{\mu})} \MCb = 0,
\eeq
the generalization of \eqref{Lie of MC = 0}, along with the fact that $\algcon{\ellsec{\mu}} \MCb=\algcon{\alginjo(\ellsec{\mu})} \MCb = \ellsec{\mu}$. We have also used $\alghatdTM$ to refer to the de Rham differential on $\fldspalg$, which should be regarded as the field variational derivative. 

%
The modification in moving from $\alghatdTM$ to $\alghatd$ (as in eqn. \eqref{deltahatalpha}) proves essential for accomplishing our goal of realizing integrable charges for local symmetries. Often, one focusses on equivariant functionals $F: \fldspalg \rightarrow \mathbb{R}$ which satisfy $F[\varphi^* \Phi] = \varphi^* F[\Phi]$ (or equivalently  the Lie derivative along $\algvec{\ellsec{\mu}}$ vanishes). However, this is rarely the case, as even the classical action is not equivariant (as in eq. \eqref{diffactionLagr} for example), while in specific gauge theories, it is not true for gauge transformations either (as in eq. \eqref{Failure of equivariance} in Chern-Simons). Part of the motivation of the extended phase space is that functionals which fail to be equivariant in the non-extended sense can be interpreted as equivariant in the extended (or configuration algebroid) sense: $\algLie{-\alginjo(\ellsec{\mu})}F=0$. This is the subject of the following section.

\section{The Extended Phase Space of the Configuration Algebroid}\label{sec: General Proof}

We now move to the main result of this paper, namely a proof of the existence of integrable charges for any local symmetry. 
In this paper, we will confine our discussion to the classical scenario, in which a symplectic structure is encoded in a Lagrangian density. 
The first important modification to the analysis of gauge theories predicated by the configuration algebroid comes in the specification of such a  Lagrangian. As we have stressed, the degrees of freedom relevant to a gauge theory are sections of Atiyah Lie algebroids and their associated vector bundles. To encode a theory in a manner that is consistent with this observation, we are led to define a Lagrangian as a differential form in the exterior algebra of some $A$. In particular, for a theory living on a $d$-dimensional spacetime, the Lagrangian is taken to be a $d$-form, $\Lagr \in \Omega^d(A) = \Omega^0(\algalg;\Omega^d(A))$.\footnote{Here we have stressed that the Lagrangian should also be regarded as a $0$-form in the configuration algebroid taking values in $\Omega^d(A)$, which is regarded as a bundle associated with $\algalg$. We should also note that where derivatives $\td$ appear in the usual Lagrangian, $\hatd$ should now replace it. This appears, for example, below in eq. \eqref{deltahatLagr}.\label{footie14}} Here, we should be careful to note that, viewed as a form in $\Omega^0(\algalg;\Omega^d(A))$, the Lagrangian is allowed to vary over different gauge inequivalent values for the possible field configurations included in $\fldspalg$. 

Replacing the familiar spacetime top form Lagrangian with a Lagrangian which is a differential form in the complex of an Atiyah Lie algebroid may seem odd; however, it is useful in that it naturally allows for the physics of submanifolds or subregions to be absorbed into the formulation of the theory. For example, in Section \ref{Examples}, we demonstrate how, in the case of Chern-Simons theory, the extended Lagrangian form automatically includes terms relevant to submanifolds. 

Given such a  Lagrangian $\Lagr$, we can now provide a mathematical language that will subsume all of the analysis outlined in Section \ref{sec:conventional}. To start, we obtain the extended presymplectic potential by taking the extended variation of the Lagrangian. This is just an application of the general formula \eqref{deltahatalpha}, yielding 
\beq\label{deltahatLagr}
\alghatd\Lagr =  \mathcal{E}_I \alghatdTM \Phi^I+ \hatd\theta 
+ {\text L}_{ \algvec{\ellsec{\MCb}}}\Lagr 
=
\mathcal{E}_I \alghatdTM \Phi^I + \hatcon{\MCb} \hatd \Lagr + \hatd\left(\theta + \hatcon{\MCb} \Lagr\right)
\eeq
where the $\algEnd{\mathbf{E}}(\MCb)$ appearing in \eqref{deltahatalpha} reduces in this case to simply the Lie derivative on $T\fldspalg$ (see footnote \ref{footie14}). The term $\hatcon{\MCb} \hatd \Lagr$ is often zero (for example, in Chern-Simons theory, $\hatd \Lagr$ is proportional to a characteristic class, is horizontal and thus vanishes.)
In the above expression, $\theta$ is the non-extended presymplectic potential and
\beq
	\theta^{ext.} = \theta + \hatcon{\MCb} \Lagr \in \Omega^1(\algalg; \Omega^{d-1}(A))
\eeq	
which is reminiscent of the extended symplectic forms appearing in
 \eqref{Sec 2 Extended Potential} and \eqref{sec 2 extended potential 2}, if we realize that $\MCb$ subsumes the roles played by \emph{both} $\un{\chi}$ and $\lambda$. We will see this correspondence in detail in the next section. Thus, extending the phase space with respect to the algebroid $A$ has the desired effect of extending the phase space with respect to diffeomorphisms \emph{and} gauge transformations, simultaneously. This exemplifies why the algebroid is an effective construction: it presents gauge transformations on the same footing as spacetime diffeomorphisms and thus both appear via Lie derivatives on the algebroid in a way that is fully gauge covariant.

What we would now like to show is that in the algebroid language we have simply the pair of equations
\beq \label{Covariance of L and Theta 2}
	\algLie{-\alginjo(\ellsec{\mu})} \Lagr = 0, \qquad \algLie{-\alginjo(\ellsec{\mu})} \theta = 0.
\eeq
%
This is where the utility of the extended phase space truly begins to present itself. As was originally noted in \cite{Ciambelli:2021nmv}, the extended phase space \emph{absorbs} the anomaly into the definition of the extended Lie derivative. To be precise:
\beq
	\algLie{-\alginjo({\ellsec{\mu}})} \Lagr = \algcon{-\alginjo({\ellsec{\mu}})} \alghatd \Lagr = {\text I}_{\algvec{\ellsec{\mu}}} \alghatdTM \Lagr - {\text L}_{ \algvec{\ellsec{\mu}}} \Lagr =0
\eeq
where we have made use again of an instance of eq. \eqref{deltahatalpha}. Similarly
\begin{flalign}
	\algLie{-\alginjo({\ellsec{\mu}})} \theta &= \alghatd \algcon{-\alginjo({\ellsec{\mu}})} \theta + \algcon{-\alginjo({\ellsec{\mu}})} \alghatd \theta
\\
	&= \alghatd {\text I}_{\algvec{\ellsec{\mu}}} \theta + {\text I}_{\algvec{\ellsec{\mu}}} \alghatd \theta - \algcon{\ellsec{\mu}} \alghatd \theta
\\
	&= \cancel{\alghatdTM {\text I}_{\algvec{\ellsec{\mu}}} \theta }
	+ {\text L}_{ \algvec{\ellsec{\MCb}}} {\text I}_{\algvec{\ellsec{\mu}}} \theta 
-	 {\text I}_{ \algvec{[\ellsec{\mu},\ellsec{\MCb}]_{\ellalg}}}  \theta 
	+ \cancel{{\text I}_{\algvec{\ellsec{\mu}}} \alghatdTM \theta }
	+ {\text I}_{\algvec{\ellsec{\mu}}} {\text L}_{ \algvec{\ellsec{\MCb}}}\theta
	 - \bcancel{\algcon{\ellsec{\mu}} \alghatdTM \theta }
	 - \cancel{\algcon{\ellsec{\mu}} {\text L}_{ \algvec{\ellsec{\MCb}}}\theta}
\\
	&= 
	{\text L}_{ \algvec{\ellsec{\MCb}}} {\text I}_{\algvec{\ellsec{\mu}}} \theta 
+	 {\text I}_{ [\algvec{\ellsec{\mu}},\algvec{\ellsec{\MCb}}]_{T\fldspalg}}  \theta
	+ {\text I}_{\algvec{\ellsec{\mu}}} {\text L}_{ \algvec{\ellsec{\MCb}}}\theta
	= 0
\end{flalign}
In the second line, we have made use of eq. \eqref{Structure Maps of Configuration Algebroid} and $\algcon{\ellsec{\mu}} \theta = 0$. In the third line, we have used the fact that $\alghatd \alpha = \alghatdTM \alpha^a \otimes \bf{e}_a + {\text L}_{ \algvec{\ellsec{\MCb}}} \alpha$ for all $\alpha \in \Omega^{(p,0)}(\algalg;\Omega^q(A))$, which includes $\theta$ and $\algcon{\algvec{\ellsec{\mu}}}\theta$. Finally, in the last line we have used the fact that $\algvec{}$ represents $\ellalg$ on $T\fldspalg$.

This brings us to the main point of the paper. The pair of equations \eqref{Covariance of L and Theta 2} imply that the symmetry generators $-\alginjo(\ellsec{\mu}) \in \algalg$ induce integrable charges in the context of the configuration algebroid. To see that this is the case consider
\beq
	\algLie{-\alginjo(\ellsec{\mu})} \theta^{ext.} = \algLie{-\alginjo(\ellsec{\mu})} \theta + \hatcon{\MCb} \algLie{-\alginjo(\ellsec{\mu})} \Lagr + \hatcon{\algLie{-\alginjo(\ellsec{\mu})} \MCb} \Lagr.
\eeq
By \eqref{Covariance of L and Theta 2}, the first two terms are equal to zero. Thus, we find
\beq \label{Lie derivative of Theta ext}
	\algLie{-\alginjo(\ellsec{\mu})} \theta^{ext.} = \hatcon{\algLie{-\alginjo(\ellsec{\mu})}\MCb} \Lagr
\eeq
But \eqref{Lie derivative of Theta ext} is immediately zero by \eqref{Lie MC = 0}! Thus, we have shown that\footnote{Here, $\omega^{ext.} \equiv \alghatd \theta^{ext.}$ is the symplectic form density.}
\beq \label{Integrable charges}
	0 = \algLie{-\alginjo(\ellsec{\mu})} \theta^{ext.} = \alghatd \algcon{-\alginjo(\ellsec{\mu})} \theta^{ext.} + \algcon{-\alginjo(\ellsec{\mu})} \alghatd \theta^{ext.} = \alghatd \algcurrdens{\ellsec{\mu}} + \algcon{-\alginjo(\ellsec{\mu})} \omega^{ext.}.
\eeq
Here, we have defined $J: \ellalg \rightarrow \Omega^{d-1}(A)$, as the Noether current densities  associated with symmetry generators $\ellsec{\mu} \in \ellalg$. As is indicated by \eqref{Integrable charges}, $J$ is a \emph{moment map}. Notice, moreover, that $J = -\alginjo^* \theta^{ext.}$, or explicitly
\beq \label{current extension}
	\algcurrdens{\ellsec{\mu}} :=\algcon{-\alginjo(\ellsec{\mu})} \theta^{ext.}
	= {\text I}_{\algvec{\ellsec{\mu}}} \theta - \hatcon{\ellsec{\mu}} \Lagr.
\eeq
A moment map which can be written as the pullback of an invariant presymplectic potential can be shown to induce a \emph{moment morphism} \cite{guillemin1984normal,kostant1970quantization,guillemin1990symplectic
} $\algchg{\ellsec{\mu}}=\int\varphi^*(\algcurrdens{\ellsec{\mu}})$  in the respect that
\beq \label{No central extension}
	 \algcon{-\alginjo(\ellsec{\mu})}\algcon{-\alginjo(\ellsec{\nu})}\omega^{ext.} - \algcurrdens{[\ellsec{\mu},\ellsec{\nu}]_{\ellalg}} = 0.
\eeq
Equation \eqref{No central extension} implies by integration that the algebra of charges is isomorphic to the algebra of symmetry generators, and therefore possesses \emph{no central extension}. We should mention that \eqref{No central extension} is not the end of the story as it pertains to the central extension. Strictly speaking we have only shown that \eqref{No central extension} holds in a local sense, and in particular we have not shown that $\theta^{ext.}$ is a globally defined object. The presence of a central extension is understood to be a cohomological obstruction which describes the global topology of the configuration space of a theory \cite{brown1986central,barnich2002covariant,brown1982cohomology,TUYNMAN1987207}. In \cite{Klinger:2023abc} we will provide a more complete appraisal of when the central extension of a charge algebra truly vanishes in a global sense.

We have therefore demonstrated that by extending the phase space via the configuration algebroid, one obtains a proper representation of the algebra. At no point in our argument have we appealed to  equations of motion or the constraints, or imposed extraneous conditions such as boundary conditions. Our result is fully general for any section of $\ellalg$, and requires no restriction to sections that are field independent. All of this is to say that our conclusion appears to be a statement about the \emph{geometry} of the configuration algebroid itself, rather than about any particular theory or clever system of constraints.

\section{Application to Chern-Simons}\label{Examples}

Let us now provide some clarity on the very abstract presentation of the previous sections by revisiting $3d$ Chern-Simons theory that we considered in more standard formalism in Section \ref{sec:conventional}. In an appendix, we discuss a second example, Einstein-Yang-Mills theory. The results can be easily generalized to an arbitrary covariant theory.
In doing so we will demonstrate that the configuration algebroid construction not only automatically reproduces the charge algebra described in Section \ref{sec:conventional}, but we also absorb many other physically relevant features into our description of the theory.

The first step in analyzing Chern-Simons theory in the context of the configuration algebroid is to promote the configuration space utilized in Section \ref{sec:conventional} to the extended configuration space. As we have introduced in Section \ref{sec:confalg}, moving from the conventional configuration space, which in the case of Chern-Simons theory is simply the set of gauge fields, to the extended configuration space consists of two steps. First, we must promote the space of gauge fields to the space of Lie algebroid connection reforms. And second, we must append to this space the set of Lie algebroid morphisms, corresponding to diffeomorphisms and gauge transformations as established in Section \ref{sec:Primer}. Thus for Chern-Simons theory we take locally
\beq
\pbalg = \{(\varphi,\omega) \;|\; \varphi \in \morphalg, \omega \in \Omega^1(A;L)\},
\eeq
where $\omega$ satisfies the properties required of a connection reform. This gives rise to the Atiyah Lie algebroid $\algalg$.

Having moved from the space of gauge fields to the extended configuration space of Lie algebroid connection reforms and their gauge orbits, we must now perform an analogous promotion of the conventional spacetime Lagrangian \eqref{CS Spacetime Lagrangian} into a $3$-form in the exterior algebra of a Lie algebroid, $A$.  The appropriate Chern-Simons Lagrangian form is given by:
\beq\label{CS Algebroid}
    \Lagr[\omega] = \frac{k}{4\pi} tr\left(\omega \wedge \hatd \omega + \frac{1}{3} \omega \wedge [\omega \wedge \omega]_L\right).
\eeq
In \cite{Jia:2023tki} it is shown that \eqref{CS Algebroid} is indeed a bona-fide Chern-Simons form in the respect that $\hatd \Lagr[\omega]$ gives rise to the second Chern class suitably generalized to the algebroid cohomology. Explicitly, $\hatd \Lagr[\omega] = \frac{k}{4\pi}\text{ch}_2(\Omega) = \frac{k}{4\pi} tr\left(\Omega \wedge \Omega\right)$.\footnote{More generally, in \cite{Jia:2023tki} it is shown in our context how the Chern-Weil theory connecting characteristic classes quantifying topological obstruction and cohomology classes is carried into the Lie algebroid context. Using this machinery, it is possible to represent arbitrary characteristic classes, including the Chern classes of arbitrary dimension, and their associated Chern-Simons forms in the algebroid language. For a mathematically rigorous approach to this subject, see \cite{fernandes2002lie}} The Chern class $\text{ch}_2(\Omega)$ is a  horizontal form because the curvature of a Lie algebroid connection is purely horizontal \cite{Ciambelli:2021ujl}. In a three-dimensional Chern-Simons theory we consider the algebroid $A$ as being based on a three-dimensional manifold, $M$. Hence, the top horizontal form degree in the exterior algebra of $A$ is three, and the four form $\text{ch}_2(\Omega)$ is identically zero. 

Lifting the Lagrangian from a spacetime top form to a Lie algebroid $3$-form has the effect of encoding the additional degrees of freedom associated with subregions of spacetime which are necessary to complete the extended phase space. An immediate indication of this fact is that the Chern-Simons action is not purely a horizontal form in the exterior algebra of $A$ \cite{Jia:2023tki,Prabhu:2015vua}. This means that the contribution to the currents from the second term in \eqref{current extension} will include an explicit correction to the gauge charges. To illustrate these degrees of freedom more clearly, and to connect our subsequent results more directly with conventional literature, we will, as in previous sections describe  $(A,\omega)$ in the consistent splitting (as in  \cite{Jia:2023tki}, this can also be thought of as working in the equivalent trivial Lie algebroid $(A_{\tau}, \omega_{\tau})$.) Thus, we will simply associate $\omega=b-\varpi$, where $b \in \Omega^1(M,L)$ is the gauge field associated with the connection $\omega \in \Omega^1(A,L)$, and $\varpi \in \Omega^1(L,L)$ is the Maurer-Cartan form of the adjoint bundle $L$. 

We then have
\beq\label{Trivialization of CS Algebroid}
\Lagr[\omega] = \Lagr[b] + \frac{k}{4\pi} tr\left(-\varpi \wedge db - \frac{1}{2}b \wedge [\varpi \wedge \varpi] + \frac{1}{6}\varpi \wedge [\varpi \wedge \varpi]\right).
\eeq
The first term on the right hand side of \eqref{Trivialization of CS Algebroid} is the spacetime Lagrangian of the Chern-Simons theory as written in \eqref{CS Spacetime Lagrangian}. The remaining terms play a crucial role in governing the topological characteristics of the Chern-Simons form in relation to the role of Chern-Simons theory in quantifying anomalies \cite{Jia:2023tki}. These terms can also be interpreted as quantifying physics on lower dimensional subregions of the three dimensional bulk. In other words, writing the Lagrangian as a three-form in the exterior algebra of the algebroid $A$ lends itself naturally to the idea of a so-called ``extended action" relevant for specifying the dynamics of edge modes as well as the gluing of subregions \cite{Fliss:2017wop,balasubramanian2017multi,Geiller:2019bti,balasubramanian2018entanglement,witten1988quantum,Dong:2008ft,Wong:2017pdm,Donnelly:2018ppr,Blommaert:2018oue}. In this vein, it would be particularly interesting to directly investigate how these added terms help in clarifying the entanglement properties of the Chern-Simons theory through the extended phase space formalism introduced in this paper. We leave this question to future work. 

We can now follow the procedure outlined in Section \ref{sec: General Proof} to derive the algebra of integrable charges associated with the local symmetries of the Chern-Simons theory. Firstly, we compute following \eqref{deltahatLagr}
\beq
    \alghatd\Lagr[\omega] = \frac{k}{4\pi} tr\left( \alghatdTM\omega \wedge \Omega\right) + \hatd\left(\frac{k}{4\pi} tr(\alghatdTM \omega \wedge \omega) + \hatcon{\MCb}\Lagr[\omega] \right)
\eeq
from which we can recognize the standard equation of motion, $\Omega \; \hat{=} \; 0$, the non-extended presymplectic potential $\theta = tr(\alghatdTM\omega \wedge \omega)$, and the extended presymplectic potential $\theta^{ext.} = \theta + \hatcon{\MCb} \Lagr[\omega]$. Notice that this consolidates the extensions discussed in Section \ref{sec:conventional} with respect to diffeomorphisms \eqref{Sec 2 Extended Potential} and gauge transformations \eqref{sec 2 extended potential 2} into a single extension with $\MCb$ subsuming the roles played by $\underline{\chi}$ and $\lambda$. This form of $\theta^{ext.}$ is more geometrically meaningful than that discussed in \eqref{sec 2 extended potential 2}, since we can now explicitly encode the extension with respect to gauge transformations as a contraction with the Lagrangian in precisely the same way as the diffeomorphism extension, due to the fact that elements of the gauge algebra and diffeomorphisms are combined into a single section of the algebroid $A$. 

Based on the conclusion of Section \ref{sec: General Proof}, Chern-Simons theory should give rise to a fully integrable charge algebra with respect to both diffeomorphism and gauge symmetries when regarded from the perspective of the configuration algebroid. Let us now compute these charges explicitly. Using \eqref{current extension} we identify the currents:
\beq \label{Noether Charge}
    \algcurrdens{\ellsec{\gmu}} = \algcon{-\alginjo(\ellsec{\gmu})} \theta^{ext.} = \algcon{\algvec{\ellsec{\gmu}}} \theta - \hatcon{\ellsec{\gmu}} \hatcon{\MCb} \Lagr \, \hat{=} \, \frac{k}{4\pi} \hatd \left( tr(\hatcon{\ellsec{\gmu}} \omega \; \omega)\right)
\eeq 
 Thus, when computed on shell, the current form is a total $\hatd$ coboundary, meaning that the charges of the theory will be codimension-$2$ objects. 

To see more clearly that \eqref{Noether Charge} reproduces exactly the results derived in Section \ref{sec:conventional}, let us evaluate it in the consistent splitting
\begin{flalign}
    \algcurrdens{\ellsec{\gmu}} &= \frac{k}{4\pi} \hatd \left( tr\left(\hatcon{\underline{\xi}\oplus -\gmu}(b \oplus -\varpi) 
    \; (b \oplus -\varpi)\right)\right) \\
    &= \frac{k}{4\pi} \td \left(tr\left((i_{\underline{\xi}} b + \gmu) \; (b \oplus -\varpi)\right)\right) \\
    &= \frac{k}{4\pi} \td \left(tr\left((i_{\underline{\xi}} b + \gmu) \; b \right) + ... \right) 
\end{flalign}
Here $...$ denotes terms which are not codimension-$1$ in the exterior algebra of the base manifold $M$, and therefore will not contribute to an integral over a hypersurface in $M$. Given such a hypersurface, $\Sigma \subset M$ possessing a boundary $\partial \Sigma$ which is a corner in $M$, we can write the charge associated with a local symmetry as:
\beq\label{CS Charges}
\algchg{\ellsec{\gmu}} 
\equiv \int_{\Sigma} \varphi^* \algcurrdens{\ellsec{\gmu}} 
= \frac{k}{4\pi} \int_{\Sigma} \varphi^*\td \left( tr\left((i_{\underline{\xi}} b + \gmu) b \right)\right)
 = \frac{k}{4\pi} \int_{\partial \Sigma}\varphi^* tr\Big((i_{\underline{\xi}} b) b\Big) + \frac{k}{4\pi} \int_{\partial \Sigma} \varphi^* tr(\gmu b)
\eeq
In Appendix C, we explain how an embedding map can be related to an algebroid morphism; this is why we have written the above in terms of $\varphi$. 
These are precisely the Hamiltonian charges computed in Section \ref{sec:conventional}, with the first term on the right hand side of \eqref{CS Charges} corresponding to the diffeomorphism charge and the second term corresponding to the gauge charge. From this point it is straightforward to show that we also realize the charge algebra identified in \eqref{Charge Algebra}. These results can now be organized into a single expression: $\{\algchg{\ellsec{\gmu}},\algchg{\ellsec{\gnu}}\} = \algchg{[\ellsec{\gmu},\ellsec{\gnu}]_{\ellalg}}$.

In this section we have illustrated how the charge algebra of Chern-Simons theory is seamlessly encoded through the extended phase space of the configuration algebroid. Moreover, in dissecting the procedure by which the standard covariant phase space is promoted to the extended phase space, we have understood in a very explicit way what new degrees of freedom were needed to  make these results possible.

\section{Discussion}

In this paper, we have presented an approach to analyzing the extended phase space which is amenable to theories with generic combinations of internal and spacetime symmetries. Our approach centers around a new construction called the \emph{configuration algebroid}. The configuration algebroid arises as the Atiyah Lie algebroid derived from a principal bundle of field configurations whose structure group(oid) is the set of spacetime local gauge transformations \emph{and} diffeomorphisms. In Section \ref{sec: General Proof} we demonstrated that the action of local gauge transformations and diffeomorphisms on the (pre)symplectic geometry derived from any Lagrangian theory 
is {Hamiltonian} when considered in the context of the extended phase space of the appropriate configuration algebroid. That is, the algebr(oid) of symmetries formed from the combination of gauge transformations and diffeomorphisms is faithfully encoded in the set of currents by means of an equivariant moment map which is directly related to the extended pre-sympectic potential. This result was proven in complete generality without appealing to any specific theory, equations of motion, or constraints, and should therefore be considered a geometric property of the configuration algebroid. In this respect, we argue that the configuration algebroid provides a satisfactory accounting of the relevant degrees of freedom needed to properly analyze generic gauge theories, including gravity. 

In Section \ref{Examples}, we re-analyzed $3d$ Chern-Simons theory and in Appendix \ref{EYMApp} Einstein-Yang-Mills theory using the configuration algebroid to illustrate the usefulness of our approach and connect with existing literature. In particular, we observed that the extended Lagrangian form required for analysis in the context of the configuration algebroid naturally incorporates terms quantifying the physics of subregions into the setting of the theory. The addition of these terms, which has been introduced as a computational device in previous literature, now has a clear geometric origin. We anticipate that the presence of these terms will make the configuration algebroid a natural tool for analyzing the entanglement properties of gauge theories. Finally, as expected, the Noether currents computed for both of the specific theories we considered are realized as degree $d-1$ forms that are exact with respect to the coboundary operator $\hatd$. Hence, the charges of the theories, computed in \eqref{CS Charges} and \eqref{EYM Charges}, localize on subspaces of codimension-$2$ (corners) in a $d$-dimensional bulk.

This work represents an important step forward in the approach to analyzing the symplectic geometry of gauge theories; however there remain many unresolved questions. The most immediate question concerns the procedure of lifting the \emph{pre}symplectic structure described in this paper to a manifestly symplectic one, in other words the problem of symplectic reduction \cite{Marsden:1974dsb,Bottacin2008AMR,MR893479}. In physics contexts, this procedure entails distinguishing the set of charges which are physical as opposed to pure gauge. The investigation of this problem for theories with only diffeomorphisms led to the discovery of the universal corner symmetry group (UCS), which is the maximal non-trivial (physical) symmetry algebra at isolated corners \cite{Ciambelli:2021vnn,Ciambelli:2022cfr,Freidel:2021cjp}. We anticipate that the configuration algebroid will play a central role in solving this problem for general gauge theories, as we consider the analog of the UCS in cases where internal symmetries are present in addition to diffeomorphisms. The configuration algebroid is particularly well suited for such a task because it is agnostic to the classical theory in question, rather concentrating on the geometric aspects of the symmetries which are present. In this way, we anticipate that the configuration algebroid can be especially useful if it is paired with a mechanism for constructing theories which begin with the symplectic structure as its defining feature, as opposed to the present approach which starts from a classical Lagrangian. Such an approach, which for now we will refer to as the \emph{Hamiltonian Algebroid}, is the topic of a forthcoming paper \cite{Klinger:2023abc}. 

Another interesting problem which we have addressed at least in part in this paper concerns the existence of a central extension to the algebra of charges associated with gauge symmetries. It is well known that a centrally extended charge algebra is a hallmark of theories with \emph{anomalies}.\footnote{In the companion paper \cite{Jia:2023tki}, we have addressed the analysis of anomalies from the Lie algebroid perspective. In future work we plan to combine this analysis with the extended phase space.} 
It turns out that this problem, too, is best addressed in the context of the Hamiltonian Algebroid. As we have alluded to, whether the Poisson algebra of Hamiltonian functions associated with a given symplectic form is centrally extended is related to the existence of an equivariant lifting of said symplectic form, and, by extension, the presence of a globally well defined equivariant moment map \cite{kostant1970quantization,atiyah1984moment,guillemin1990symplectic}. In \cite{Klinger:2023abc}, we will demonstrate that the obstruction to equivariant lifting is quantified by the curvature of a would-be morphism between the configuration algebroid and the Hamiltonian Algebroid, which, moreover, is equivalent to the central extension of the charge algebra. This will provide a more detailed account of when one should or should not expect a central extension from a manifestly (pre)quantum perspective that has no dependence on the existence of an underlying classical theory.

In this paper we have worked with the conceit of a classical spacetime, and subsequently have stressed the importance of subregions in dictating the physical degrees of freedom. However, in \cite{Ciambelli:2022cfr} it was argued that the physical degrees of freedom associated with diffeomorphism-invariant theories should really be considered as arising from a geometric structure that, while recognizable as a subalgebra of the diffeomorphisms of a $d$-dimensional spacetime, exists independently without the need to specify a classical spacetime at all. Again, this idea is motivated by the notion that classical physics should emerge from quantum physics, and not the other way around. The Hamiltonian algebroid will be an important ingredient in achieving these goals.


\paragraph{Acknowledgments}

We thank Luca Ciambelli, Weizhen Jia and Michael Stone for discussions. 
This work was supported by the U.S. Department of Energy under contract DE-SC0015655. 

\appendix
\renewcommand{\theequation}{\thesection.\arabic{equation}}
\setcounter{equation}{0}

\section{Notation Conventions}\label{AppNot}

Throughout the text, we use an underline notation, such as $\un\xi$, for a space-time vector field. The Lie derivative of a tensor along a vector field $\un\xi$ is denoted ${\cal L}_{\un\xi}$, while the contraction of $\un\xi$ with a differential form is written $i_{\un\xi}$. The Cartan magic formula is then written ${\cal L}_{\un\xi}=i_{\un\xi}\td+\td i_{\un\xi}$.

Moving into a Lie algebroid we use $\mX$ to denote a section of $A$, and $\gmu$ to denote a section of $L$. The Maurer-Cartan form for $L$ is written $\varpi$. The algebroid Lie derivative is denoted $\hatLie{\mX}$, while contraction of $\mX$ with a section of $\Omega(A;E)$ is written $\hatcon{\mX}$. Sections of a vector bundle $E$ associated to $A$ are denoted by $\un{\psi}$. The Lie algebroid representation associated to $E$ is given by $\Aconn{E}$, and can be used to define the algebroid coboundary operator $\hatd$. We can then carry the Cartan formula into the algebroid context as $\hatLie{\mX} = \hatd \hatcon{\mX} + \hatcon{\mX} \hatd$.

We refer to $\fldspalg$ as the naive field space which consists simply of all possible field configurations for a given theory. We denote a generic section of $\text{T}\fldspalg$ by $\algvec{}$, while we denote a section of $\text{T}\fldspalg$ which generates a diffeomoprhism by $\algvec{\un{\xi}}$, and a section of $\text{T}\fldspalg$ that generates a gauge transformation by $\algvec{\gmu}$. We use $\fsLie{}$ and $\fscon{}$ to refer to the naive field space Lie derivative and contraction, respectively. On $\fldspalg$ there is a coboundary operator $\alghatdTM$ arising from the variation derivative. We may then carry the Cartan formula to this context as $\fsLie{} = \alghatdTM \fscon{} + \fscon{} \alghatdTM$.

The configuration algebroid is denoted by $\algalg$, and possesses an isotropy bundle $\ellalg$ which is a bundle of Lie algebroids over $\fldspalg$. We denote a section of $\algalg$ by $\algsec{X}$, and a section of $\ellalg$ by $\ellsec{\mu}$ which can be identified as a field dependent section of an algebroid $A$. The Maurer-Cartan form for $\ellalg$ is written $\MCb$. Carrying over notation from the naive field space, we denote by $\algvec{\ellsec{\mu}}$ the generator of local diffeomorphisms and gauge transformations in $\text{T}\fldspalg$. We use $\algLie{\algsec{X}}$ and $\algcon{\algsec{X}}$ to refer to the the Lie derivative and contraction in the configuration algebroid. The configuration algebroid possesses a coboundary operator $\alghatd$ in analogy with $\hatd$. We may therefore write the Cartan formula in $\algalg$ as $\algLie{\algsec{X}} = \alghatd \algcon{\algsec{X}} + \algcon{\algsec{X}} \alghatd$.

\section{Einstein-Yang-Mills theory}\label{EYMApp}

As another classic example, we now illustrate how to apply the configuration algebroid to Einstein-Yang-Mills theory. The analysis is closely related to that presented in \cite{Prabhu:2015vua}, in which the Lagrangian is analyzed from the perspective of a principal bundle. The reason for the similarity in this case is that the Einstein-Yang-Mills action is horizontal and gauge covariant, in contrast to the Chern-Simons action. In this way the symplectic geometry of the Einstein-Yang-Mills theory is only extended with respect to diffeomorphisms, but not internal gauge symmetries. 

We work in the first order formalism. The internal symmetry group is given by $G=SO(1,d-1) \times K$, which we identify as the Lorentz group times a gauge group $K$. Using this structure group we define the principal $G$-bundle, $P$. To formulate the Einstein-Yang-Mills theory in the extended phase space, we consider the Atiyah Lie algebroid $A$ derived from this principal bundle. This Lie algebroid has an isotropy bundle which splits into a direct sum $L= P \times_{SO(1,d-1)} \mathfrak{so}(1,d-1) \oplus P \times_{K} \mathfrak{k}$. Hereafter we will refer to the factors of this algebra as $L_L$ and $L_K$, respectively.

The Lie algebroid $A$ comes equipped with a connection reform $\omega \in \Omega^1(A;L)$, which splits as
\beq
\omega = \omega_L \oplus \omega_K
\eeq
Here we have introduced $\omega_L \in \Omega^1(A;L_L)$ and $\omega_K \in \Omega^1(A;L_K)$, which are the projections of the connection reform down to the isotropy bundles of the Lorentz and gauge groups, respectively. The connection reform $\omega$ has a curvature which is defined in the usual way, and further splits as
\beq
\Omega = \hatd \omega + \frac{1}{2}[\omega \wedge \omega]_L = \hatd \omega_L + \frac{1}{2}[\omega_L \wedge \omega_L]_{L_L} \oplus \hatd \omega_K + \frac{1}{2}[\omega_K \wedge \omega_K]_{L_K} \equiv \Omega_L \oplus \Omega_K.
\eeq
Here $\Omega_L \in \Omega^2(A;L_L)$ and $\Omega_K \in \Omega^2(A;L_K)$. Both $\Omega_L$ and $\Omega_K$ are \emph{horizontal} forms, as discussed in \cite{Ciambelli:2021ujl}. 

To complete the formulation of the Einstein Yang-Mills theory, we must also specify an additional dynamical field called a \emph{solder form}. A solder form is a map, $\hat{e}: A \rightarrow \mathcal{E}$, where here $\mathcal{E}$ is a $d$-dimensional vector bundle associated to the Lorentz group. Like the curvatures, the solder form is a horizontal form, $\hat{e} \in \Omega^1(H;\mathcal{E})$. Given a basis of sections for $\mathcal{E}$, $\{\underline{v}_a\}_{a = 1}^d$, we can therefore write
\beq
\hat{e} = \hat{e}^a_{\underline{\alpha}} E^{\underline{\alpha}} \otimes \underline{v}_a.
\eeq
The role of the solder form is to encode the $L_L$ invariant inner product $\eta: \mathcal{E} \times \mathcal{E} \rightarrow \mathbb{R}$. That is, given a linear representation $v_{\mathcal{E}}: L \rightarrow \text{End}(\mathcal{E})$
\beq
	\eta(v_{\mathcal{E}}(\un{\mu})(\un{v}),v_{\mathcal{E}}(\un{\mu})(\un{w})) = \eta(\un{v},\un{w}), \; \forall \un{v},\un{w} \in \Gamma(\mathcal{E}), \; \un{\mu} \in L_L.
\eeq
One can always choose a basis for $\mathcal{E}$ such that $\eta = \eta_{ab} \un{v}^a \otimes \un{v}^b$. Viewing $\hat{e}: H \rightarrow \mathcal{E}$ as a map, we can also form the pullback $\hat{g} = \hat{e}^*\eta: H \times H \rightarrow \mathbb{R}$,
\beq \label{Metric in Algebroid}
	\hat{g} = \eta_{ab} \hat{e}^a \otimes \hat{e}^b.
\eeq 
Once we move into the consistent splitting of the algebroid we will see that $\hat{g}$ may be interpreted as a spacetime metric. 

The solder form is required to be invertible, and thus we also obtain a co-solder form $\underline{\hat{e}}: \mathcal{E} \rightarrow A$ with the property that
\beq
\hat{e} \circ \underline{\hat{e}}(\underline{v}) = \underline{v}, \;\;\; \forall {\underline{v} \in \Gamma(\mathcal{E})}.
\eeq
The solder form and its co-solder form should be regarded as playing the roles of a frame and co-frame in the standard first order formalism. In particular, the solder form defines a covariant top form in the exterior algebra of the horizontal distribution:
\beq
\text{Vol}_H \equiv \wedge_{a = 1}^d \hat{e}^a = \epsilon_{a_1 ... a_d} \hat{e}^{a_1} \wedge ... \wedge \hat{e}^{a_d} \in \Omega^d(H)
\eeq
This, in turn, allows us to define a notion of Hodge duality for $\Omega(H)$. In particular, $\star_H: \Omega^{k}(H) \rightarrow \Omega^{d-k}(H)$ is defined explicitly on a $k$ form $\beta \in \Omega^{k}(H)$ by
\beq
\star_{H} \beta \equiv \frac{1}{(d-k)! k!} \prod_{j = 1}^{k} \eta^{a_j c_j} \epsilon_{a_1 \cdots a_k b_1 \cdots b_{d-k}} \beta \left( \underline{\hat e}_{c_1},\cdots , \underline{\hat e}_{c_k} \right)  {\hat e}^{b_1} \wedge \cdots \wedge {\hat e}^{b_{d-k}}.
\eeq

To summarize, the extended configuration space of Einstein-Yang-Mills theory corresponds to the Lie algebroid connection reform $\omega$ and the solder form $\hat{e}$, along with the set of Lie algebroid morphisms on $A$ corresponding to diffeomorphisms of the base, gauge transformations associated with the gauge group $K$ and local Lorentz transformations:
\beq
\pbalg = \{(\varphi,\omega,\hat{e})\; | \; \varphi \in \morphalg, \omega \in \Omega^1(A;L), \hat{e} \in \Omega^1(A;\mathcal{E})\}
\eeq
The connection reform $\omega$ splits into connection reforms associated with the gauge factor and the orthogonal group, respectively. As we will show later on, in the consistent splitting these connection reforms correspond to a conventional gauge field and a spin connection, while the solder form corresponds to a conventional frame field that can be associated with a metric on $M$.

We are now ready to state the Einstein-Yang-Mills Lagrangian:
\begin{multline}\label{EYM Lagrangian}
\Lagr_{EYM} = -\frac{\epsilon_{a_1,\cdots,a_d} }{2\kappa_N (d-2)!} {\hat e}^{a_1} \wedge \cdots \wedge {\hat e}^{a_{d-2}} \wedge \left(\eta^{a_d b_d} v_{\mathcal{E}}({\Omega_L})^{a_{d-1}}{}_{b_d} - \frac{2 \Lambda}{d(d-1)} {\hat e}^{a_{d-1}} \wedge {\hat e}^{a_d} \right)\\
 + \frac{1}{4g^2} tr \left(\Omega_K \wedge \star_{H} \Omega_K\right).
\end{multline}
Notice that $\Lagr_{EYM}$ is a horizontal $d$-form since it is formulated entirely in terms of $\Omega_L$, $\Omega_K$ and $\hat{e}$, all of which are horizontal forms. 

Implementing the extended phase space construction of Section \ref{sec: General Proof}, we can construct the Noether charges of the Einstein-Hilbert theory. For any $\ellsec{\mu} \in \ellalg$, we define the Noether current $\algcurrdens{\ellsec{\mu}}$ using \eqref{current extension}:
\beq
\algcurrdens{\ellsec{\mu}} = \hatd\left(\frac{\epsilon_{a_1,\cdots,a_d}}{2\kappa_N d!} {\hat e}^{a_1} \wedge \cdots \wedge {\hat e}^{a_{d-2}} \left(\eta^{a_d b_d}\hatcon{\ellsec{\mu}} v_{\mathcal{E}}({\omega_L})^{a_{d-1}}{}_{b_d}\right) + \frac{1}{4g^2} tr(\hatcon{\ellsec{\mu}} \omega_K \star_{H}\Omega_K) \right).
\eeq
Similar to the Chern-Simons case, the generator $\ellsec{\mu}$, encodes both diffeomorphisms and internal symmetries.\footnote{At this stage it is worth noting that we are considering Einstein-Yang-Mills theory in the case of dynamical, rather than background gravity. However, our formalism is sufficiently general as to allow us to decide if we want to relegate certain fields into the background. For example, we could consider Einstein-Yang-Mills theory with background gravity by regarding only the piece of the connection reform associated with the gauge group $K$ to be part of the configuration space, and viewing the connection reform $\omega_L$ and the solder form $\hat{e}$ as background parameters. Following the prescribed analysis in this case for the Lagrangian \eqref{EYM Lagrangian}, while being careful to vary the action only with respect to the connection reform $\omega_K$, we find that the current associated with diffeomorphisms is equal to the conventional stress energy and is not a total derivative. By contrast, in the case where gravity is considered as fully dynamical, the stress energy tensor vanishes, as the gravitational stress energy term cancels with that of the matter sector as usual.}

As we did in Section \ref{Examples} we can compare $J_{\ellsec{\mu}}$ to the familiar charges defined on spacetime by moving into the consistent splitting (or, equivalently, performing a Lie algebroid trivialization). This predicates the following identifications:
\begin{flalign}\label{EYM Trivialization} 
	& \omega_K = b_K \oplus -\varpi_{K} \\
	& \Omega_K = F \\
	& \omega_L = b_L \oplus -\varpi_{L} \\
	& \Omega_L = R \\
	& \hat{e} = e \\
	& \star_H = \star
\end{flalign}
Here $\varpi_{K}$ and $\varpi_{L}$ are the Maurer-Cartan forms associated with $L_K$ and the $L_L$, respectively. As usual, $b_K \in \Omega^1(M;L_K)$ is the gauge field associated with $K$, but now we also find $b_L \in \Omega^1(M;L_L)$ which is the spin connection arising from the Lorentz factor. $F \in \Omega^2(M;L_K)$ and $R \in \Omega^2(M;L_L)$ are the gauge field strength, and Riemann curvature tensors. Finally, $e \in \Omega^1(M;\mathcal{E})$ is a frame field, and $\star$ is the space-time Hodge dual. As advertised the identification of $\hat{e}$ with the frame field $e$ implies that \eqref{Metric in Algebroid} can be interpreted as a Riemannian metric on $M$.

Using the various identifications we have described the integrable charges for Einstein Yang-Mills theory are given by:\footnote{Here we have identified $\ellsec{\mu} = \un{\xi} \oplus -(\un{\lambda} \oplus \un{\mu})$ where $\un{\xi}$, $\un{\lambda}$ and $\un{\mu}$ are field dependent elements of $TM$, $L_L$ and $L_K$, respectively. }
\begin{flalign}\label{EYM Charges}
    H_{\ellsec{\mu}} &\equiv \int_{\Sigma} \varphi^* \algcurrdens{\ellsec{\mu}} \\
&= \int_{\partial \Sigma} 
\varphi^*\left(\frac{\epsilon_{a_1,\cdots,a_d}}{2\kappa_N d!} e^{a_1} \wedge \cdots \wedge e^{a_{d-2}}\left(\eta^{a_d b_d}i_{\underline \xi} v_{\mathcal{E}}(b_L)^{a_{d-1}}{}_{b_d}\right) + \frac{1}{4g^2} tr( i_{\underline {\xi}} b_K \star F ) \right)\\
&\quad
+  \int_{\partial \Sigma}  \varphi^*\left(\frac{\epsilon_{a_1,\cdots,a_d}}{2\kappa_N d!} e^{a_1} \wedge \cdots \wedge e^{a_{d-2}}\eta^{a_d b_d}v_{\mathcal{E}}(\lambda)^{a_{d-1}}{}_{b_d}\right)
+\int_{\partial \Sigma} \frac{1}{4g^2} \varphi^*tr(\mu \star F).
\end{flalign} 

\section{The Generalization of Embedding Maps}\label{MorphEmb}

In \cite{Ciambelli:2021nmv}, the extended phase space was introduced as incorporating degrees of freedom associated with the embedding of subregions into the full configuration space of a diffeomorphism invariant theory. In the main body of this paper, we have argued that the Lie algebroid morphism \eqref{Transitive Lie Algebroid Morphism} should be interpreted as a far-reaching generalization of an embedding map, which encodes not only the physical embedding of subregions but also the choice of gauge. However, there is a subtle point here which is that a Lie algebroid morphism corresponds to a diffeomorphism whereas an embedding map needn't be a diffeomorphism. In this appendix we clarify this point by appealing to the ``tubular neighborhood theorem" which states that the embedding of any submanifold $S_k \subset M$ can be lifted to a diffeomorphism from the tubular neighborhood of $S_k$ into an open subset of $M$ containing $S_k$ \cite{bott1982differential}. 

Let $S_k \subset M$ denote a co-dimension $k$ submanifold of $M$, a $d$-dimensional manifold. A tubular neighborhood of $S_k$ in $M$ is a vector bundle $\pi: N(S_k) \rightarrow S_k$ together with a smooth map $J: N(S_k) \rightarrow M$ satisfying the following properties:
\begin{enumerate} 
	\item Let $z_{N(S_k)}: S_k \rightarrow N(S_k)$ denote the zero section of $N(S_k)$. Then,
	\beq \label{Tubular Neighborhood}
		J \circ z_{N(S_k)}: S_k \rightarrow M
	\eeq
	is an embedding of $S_k$ into $M$.
	\item There exist open subsets $U \subseteq N(S_k)$ and $V \subseteq M$ such that $\text{im}(z_{N(S_k)}) \subset U$ and $S_k \subset V$, and for which $J\rvert_{U}$ is a diffeomorphism. 
\end{enumerate}   
The tubular neighborhood theorem ensures the existence of such a tubular neighborhood for any embedded submanifold of $M$.

Now, imagine we are interested in integrating a quantity over an embedded submanifold $S_k \subset M$, as for example, in \eqref{CS Charges}. Rather than using an embedding map $\phi: S_k \rightarrow M$ in order to facilitate such an integration, we can use $\varphi: A(N(S_k)) \rightarrow A$ where $A(N(S_k))$ is an Atiyah Lie algebroid over the tubular neighborhood of $S_k$ in $M$, and the diffeomorphism defining $\varphi$ (as in \eqref{Transitive Lie Algebroid Morphism}) is given by $J: N(S_k) \rightarrow M$ as appearing above. The Lie algebroid pullback $\varphi^*: A^* \rightarrow A(N(S_k))^*$ will then naturally incorporate the embedding of $S_k$ into $M$ as a consequence of \eqref{Tubular Neighborhood}. 

To be explicit, let us restrict our attention to the case that $A = TM$, and consider the Lie algebroid morphism $\varphi: TN(S_k) \rightarrow TM$ given by the pushforward of the map $J: N(S_k) \rightarrow M$. Given a co-dimension $k$ form, $\alpha \in \Omega^{d-k}(M)$, we can write its integral over $S_k$ as
\beq \label{Integral with embedding}
	\int_{S_k} \varphi^*\alpha = \int_{S_k} J^* \alpha = \int_{S_k} J\rvert_{S_k}^* \alpha = \int_{S_k} (J \circ z_{N(S_k)})^* \alpha
\eeq
where in the final equality we have used the fact that the integral is over $S_k$, and that $J\rvert_{S_k} = J \circ z_{N(S_k)}$. By \eqref{Tubular Neighborhood}, the final equality in \eqref{Integral with embedding} is precisely the integral of the pullback of $\alpha$ by the embedding map $J \circ z_{N(S_k)}$. In the more general case where $A$ is an Atiyah Lie algebroid $\varphi$ will also include additional degrees of freedom specifying a choice of gauge, but this will not interfere with the observation made in \eqref{Integral with embedding} due to the commutativity of \eqref{Transitive Lie Algebroid Morphism}. Notice, also, that because of the chain map condition \eqref{Chain Map condition} we can write Stokes' theorem as
\beq
	\int_{S_k} \varphi^* (\td \alpha) = \int_{S_k} \td (\varphi^* \alpha) = \int_{\partial S_k} \varphi^* \alpha.
\eeq

It is worth noting that the picture we have advocated for here in which an embedding $\phi: S_k \rightarrow M$ is promoted to a diffeomorphism $J: N(S_k) \rightarrow M$ from the tubular neighborhood of the subregion into the bulk space is consistent with the derivation of the UCS as described in \cite{Ciambelli:2021vnn}. There, one considered explicitly the subgroup of diffeomorphisms of $M$ which respected the embedding of the subregion $S_k$. The basic approach was to expand tangent vectors of $S_k$ in a neighborhood of the embedding up to the point in which the associated subalgebra of diffeomorphisms closed. In light of this discussion, it is natural to interpret such a neighborhood as a tubular neighborhood of $S_k$ in $M$, which may be defined by an analogous integrability condition. Thus, understanding Lie algebroid morphisms with subregions as including diffeomorphisms with tubular neighborhoods also retains data about the emergence of the appropriate corner symmetry group, as desired. 

\providecommand{\href}[2]{#2}\begingroup\raggedright\endgroup

\end{document}